\pdfoutput=1
\documentclass[preprint,aps,prd,nofootinbib]{revtex4-2}
\usepackage{amssymb,amsmath,graphicx,color,microtype,enumerate,slashed,hyperref,xcolor,cases,verbatim,tikz,tikz-cd,multirow,booktabs,makecell}
\usepackage[figuresright]{rotating}
\usepackage[section]{placeins}
\hypersetup{linktocpage=true,colorlinks=true,linkcolor=blue,filecolor=blue,urlcolor=blue, citecolor=blue}
\graphicspath{{fig/}}
\makeatletter
\def\@hangfrom@section#1#2#3{\@hangfrom{#1#2}#3}
\def\@hangfroms@section#1#2{#1#2}
\makeatother

\begin{document}
\title{Cosmological Bell Tests with Decoherence Effects}
\author{Chon Man Sou$^{1}$}
\email{cmsou@mail.tsinghua.edu.cn}
\author{Junqi Wang$^{2, 3}$}
\email{junqi.wang@connect.ust.hk}
\author{Yi Wang$^{2, 3}$}
\email{phyw@ust.hk}
\affiliation{${}^1$Department of Physics, Tsinghua University, Beijing 100084, China}
\affiliation{${}^2$Department of Physics, The Hong Kong University of Science and Technology, \\
    Clear Water Bay, Kowloon, Hong Kong, P.R.China}
\affiliation{${}^3$The HKUST Jockey Club Institute for Advanced Study, The Hong Kong University of Science and Technology, \\
    Clear Water Bay, Kowloon, Hong Kong, P.R.China}

\begin{abstract}
The inflationary universe creates particle pairs, which are entangled in their momenta due to momentum conservation. Operators involving the momenta of the fluctuations can be rewritten into pseudo-spin operators, such as the Gour-Khanna-Mann-Revzen (GKMR) pseudo-spin. Making use of these pseudo-spin operators, cosmological Bell inequalities can be formulated. The violation of these Bell inequalities indicates the quantum nature of primordial fluctuations. 

In this work, we focus on primordial curvature perturbations. Since curvature perturbations arise from gravity, their action includes the Gibbons-Hawking-York boundary term. We clarify the role of the boundary term in selecting suitable initial conditions for linear perturbations. 

After that, we proceed to the interactions of cosmological perturbations, including the bulk and boundary interaction terms, which introduce decoherence effects. These decoherence effects change the expectation value of the Bell operator, and gradually restore the Bell inequality. We describe this process by a ``Bell test curve'', which offers a window around 5 e-folds for testing the quantum origin of cosmological perturbations. We also explore the possibility of extracting the information of the decoherence rate and the structure of primordial interactions from the Bell test curve.
\end{abstract}
\maketitle
\pagebreak
\tableofcontents
\newcommand{\ak}{\hat{a}_{\mathbf{k}}}
\newcommand{\amk}{\hat{a}_{-\mathbf{k}}}
\newcommand{\adk}{\hat{a}^{\dagger}_{\mathbf{k}}}
\newcommand{\admk}{\hat{a}^{\dagger}_{-\mathbf{k}}}
\newcommand{\bk}{\hat{b}_{\mathbf{k}}}
\newcommand{\bmk}{\hat{b}_{-\mathbf{k}}}
\newcommand{\bdk}{\hat{b}^{\dagger}_{\mathbf{k}}}
\newcommand{\bdmk}{\hat{b}^{\dagger}_{-\mathbf{k}}}

\newcommand{\vk}{\mathbf{k}}
\newcommand{\ket}[1]{\left|#1\right\rangle}
\newcommand{\bra}[1]{\left\langle#1\right|}
\newcommand{\braket}[1]{\left\langle #1 \right\rangle}
\newcommand{\sandwitch}[3]{\left\langle#1\left|#2\right|#3\right\rangle}
\newcommand{\hs}{\hspace{2pt}}
\section{Introduction}
In the paradigm of cosmic inflation~\cite{Starobinsky:1980te,Guth:1980zm,Linde:1981mu,Albrecht:1982wi,Linde:1983gd}, the large scale structure originates from quantum fluctuations that were amplified and stretched by accelerated cosmic expansion \cite{mukhanov:1981,Hawking:1982cz,Starobinsky:1982ee,Guth:1982ec,Bardeen:1983qw}. This mechanism gives a successful explanation for the observed large scale structure. However, to confirm the quantum origin of the fluctuations, we should explore ways to observe their genuine quantum effects. During inflation, the vacuum states of the quantum fluctuations evolve into two-mode squeezed states~\cite{Grishchuk:1990bj,Albrecht:1992kf}. The difference between squeezed states and semi-classical states provides a possibility to distinguish quantum and classical fluctuations. Many quantum-information concepts are introduced to investigate the squeezed state, like quantum discord~\cite{Lim:2014uea,Martin:2015qta,Kanno:2016gas}, entanglement entropy~\cite{Brandenberger:1992jh,Brandenberger:1992sr,Prokopec:1992ia,Brahma:2020zpk}, and quantum noise~\cite{Parikh:2020fhy,Parikh:2020kfh,Parikh:2020nrd}.

Based on the quantum states of primordial fluctuations, cosmological Bell tests~\cite{Campo:2005sv,Maldacena:2015bha,Choudhury:2016cso} can be constructed to show the quantum origin of the primordial fluctuations. The Bell inequality was originally formulated to study the quantum non-locality problem. It provides a testable way to distinguish between local classical hidden variables theory and quantum non-locality. In a Bell inequality, there exists a classical upper bound, and the violation of such bound means the presence of quantum nonlocality. The observation of such violation for the quantum state of the fluctuation may give us a direct answer to the question of whether the large-scale structures of our universe have a quantum origin.

To formulate cosmological Bell tests, we choose the construction of the Bell operator by GKMR pseudo-spin operators~\cite{revzen:2005,gour:2004} and find its matrix elements under the representation of field operator eigenstates (i.e., field basis). By this representation, we can naturally calculate the behavior of the expectation value of the Bell operator versus different modes (or the e-folds before/after one mode exits the horizon) under decoherence.

The construction of the Bell operator is affected by boundary terms arising from the integration-by-parts process \cite{Maldacena:2002vr}. The importance of those boundary terms is often overlooked in the literature\footnote{As discussed in \cite{Sou:2022nsd,Ning:2023ybc}, boundary terms should not be arbitrarily neglected for having the well-defined variational principle.}. In \cite{Laflamme:1993zx}, Laflamme and Matacz have already discussed quadratic boundary terms as a canonical transformation has an impact on the calculation of decoherence. It appears strange that a canonical transformation can affect physical observables, but in \cite{Grain:2019vnq}, as Grain and Vennin reveal, different canonical variables related by canonical transformation select out different vacuum states. This explains why the observable may change. We notice that there is a quadratic boundary term, which is important on superhorizon scales,
\begin{equation}
    \mathcal{L}_{\mathrm{bd}}\supset-\partial_{\tau}\left(9a^3HM^2_{pl}\zeta^2\right)\hs,
\end{equation}
emerging from doing the integration by parts to the full action of gravity and quantum fields. All the $\mathcal{L}\supset-\frac{d B(\zeta)}{d \tau}$ type boundary terms are often neglected in the literature since it does not contribute to correlator $\braket{\hat{\zeta}^n}$, and one can easily prove it with the formalism in \cite{Braglia:2024zsl}. We find that it can accelerate the squeezing process and has an impact on the expectation value of the Bell operators. We also provide a clearer explanation as to why some observables change the values under canonical transformation. As an additional example, we also show the inconsistency of calculating the squeezing parameter that appeared in literature due to choosing different conjugate momenta which are related by a quadratic boundary term.

It is very challenging to perform such cosmological Bell tests~\cite{Martin:2017zxs}. Especially, decoherence happens, which explains why our universe looks classical~\cite{Polarski:1995jg,Kiefer:2008ku}, in other words, the quantum-classical transition. The violation of Bell inequalities is fragile under decoherence. We need to consider decoherence effects in cosmological Bell tests, by calculating the expectation value of the Bell operator with the partial trace of the environment. As a result, we get the expectation value of the Bell operator as a function of the e-folds of inflation, dubbed by a ``Bell test curve''. Two pieces of information can be read off from the Bell test curve: (i) We can examine the robustness of the Bell inequality, and  (ii) extract the information of cosmological interactions from decoherence, if these cosmological Bell inequalities can be tested in the future.

To calculate the Bell test curve, we consider the gravitational interaction of single-field slow roll inflation. Both the leading bulk interaction terms~\cite{Nelson:2016kjm,Gong:2019yyz,Burgess:2022nwua} and boundary interaction terms~\cite{Sou:2022nsd,Ning:2023ybc} are considered. Since no additional fields or interactions are introduced, these effects can be seen as the lower bound of total decoherence. The Bell test curves for these two sources of decoherence have very different features. The range of the violation of Bell inequalities is shown by the Bell test curve, and for the decoherence caused by the boundary term, the Bell inequality is only slightly violated in a small range. By observing the impact on the Bell operators, it is also verified that the boundary term decoherence is indeed a physical effect, despite that the interaction structure is simpler due to the lack of time integration of interactions~\cite{Sou:2022nsd,Ning:2023ybc}.

To illustrate the idea of using the Bell test curve to probe the process of cosmological decoherence, we also provide another example of the power-law decoherence rate. In this example, in the superhorizon limit, the logarithmic Bell test curve is linearly dependent on the e-folds, and its slope and intercept can carry the information of the dominated decoherence rate which we choose a power function of the scale factor.

This paper is organized as follows. In Section~\ref{sec:set up}, we review the quantum state of the fluctuation in the form of wave functional, and the decoherence caused by gravitational nonlinearity. Subsequently, we briefly review the Bell inequality in the spin 1/2 system and then introduce the pseudo-spin operators and the construction of Bell operators for continuous variable systems. In Section~\ref{sec: effect of quadratic}, we consider the quadratic boundary term and canonical transformation and explain why some observables change when considering quadratic boundary terms. In Section~\ref{sec: expectation value}, we calculate the expectation value of the Bell operator in detail by field basis approach and verify the GKMR operator does not change under the canonical transformation from the quadratic boundary term. In Section~\ref{sec: effect of decoherence}, we introduce the concept of the Bell test curve and provide two examples. In the first example, we examine the robustness of the Bell inequality and discuss the features of the Bell test curve. In the second example, we consider the power-law decoherence rate and show the possibility of extracting information of decoherence from the Bell test curve in the superhorizon regime. The size of correction by higher loop contribution is estimated in Appendix~\ref{app:higher loop}. Section~\ref{sec:conclusion} is the conclusion and outlook.   

\section{Bell inequalities and decoherence in the sky}
\label{sec:set up}
The key observable in the Bell inequality is the expectation value of the Bell operator $\hat{\mathcal{B}} $ by $\left\langle\hat{\mathcal{B}}\right\rangle=\operatorname{Tr}\left(\hat{\rho} \hat{\mathcal{B}}\right)$. In this section, we first introduce the scalar perturbation during cosmic inflation and its decoherence. The quantum state of the fluctuation is a two-mode squeezed state. We will express this state with a wave functional and corresponding density matrix. The density matrix with the consideration of decoherence is also derived. Then we review the Bell inequality in the cosmological context. We start from the simplest Bell inequality in the spin-1/2 system and then generalize the Bell operator for the cosmological Bell inequality with the so-called pseudo-spin operators.

\subsection{The quantum state of Gaussian fluctuations}
\label{subsec:quantum state}
According to cosmological perturbation theory, the action and Lagrangian density of scalar fluctuation in $\zeta$-gauge are given by (one can see~\cite{Wang:2013zva,Baumann:2018muz} for a review)
\begin{equation}
        S \equiv\int d \tau \int d^3x  \mathcal{L}_0= \int d \tau \int d^3x \epsilon M_{pl}^2 a^4\left[\dot{\zeta}^2-\frac{1}{a^2}(\partial_i \zeta)^2\right]\hspace{2pt},
\label{equ:zeta action}
\end{equation}
where $\zeta$ is the comoving curvature perturbation, $M_{pl}$ is the Planck mass and $\epsilon\equiv-\frac{\dot{H}}{H^2}$ is the slow-roll parameter. We quantize the field $\zeta$ through canonical quantization. 

\begin{equation}
\begin{aligned}
    \hat{\zeta}_{\vk}(\tau) &\equiv f_k(\tau) \ak + f_k^*(\tau)\admk\hspace{2pt},\\
    \hat{\pi}_{\vk}(\tau) &\equiv -i\left[g_k(\tau)\ak-g_k^*(\tau)\admk\right] \hspace{2pt},
\label{equ:quantization}
\end{aligned}
\end{equation}    
where $f_k, g_k$ are mode functions, and $\ak, \admk$ are annihilation and creation operators respectively. We also use the Fourier decomposition convention $\zeta(\mathbf{x})\equiv\int \frac{d^3\mathbf{k}}{(2\pi)^3}\zeta_{\vk}e^{i\mathbf{k}\cdot\mathbf{x}}$ here, and $\pi\equiv\partial \mathcal{L}_0/\partial \zeta'$ is the conjugate momentum where prime is denoted to derivative to conformal time $\tau$. Notice that the slow-roll parameter $\epsilon$ is also a function of time. Then we write the Euler-Lagrange equation,
\begin{equation}
    \ddot{\zeta}+(3+\eta)H\dot{\zeta}+\frac{k^2}{a^2}\zeta=0 \hspace{2pt},
\end{equation}
where $\eta\equiv\frac{\dot{\epsilon}}{H\epsilon}$ is another slow-roll parameter. Applying the Fourier decomposition, then we have $\ddot{f_k}+(3+\eta)H\dot{f_k}+\frac{k^2}{a^2}f_k=0$. 
The solution of this equation in the slow-roll approximation is (also applied Bunch-Davies condition) 
\begin{equation}
f_k(\tau) = \frac{H}{M_{pl} \sqrt{4\epsilon k^3}}(1+ik\tau)e^{-ik\tau}\hspace{2pt},
\label{equ: mode of spectator}
\end{equation}
and the coefficient $g_k(\tau)$ in conjugate momentum is given by
\begin{equation}
  g_k(\tau) = i  
  \frac{M_{pl}\sqrt{k\epsilon}}{H\tau} e^{-ik\tau}\hspace{2pt}.  
\end{equation}

Starting from the the initial Bunch-Davies vacuum, the state $\ket{\Omega(\tau)}$ at some conformal time $\tau$ is given by
\begin{equation}
   \forall \vk: \hat U(\tau,\tau_0) \ak \hat U^{\dagger}(\tau,\tau_0) \left|\Omega(\tau)\right\rangle =0\hspace{2pt},
\end{equation}
where $\hat U(\tau,\tau_0)$ is the evolution operator of the free theory. 
Express the annihilation operator with $\hat{\zeta}_{\vk}$ and $\hat{\pi}_{\vk}$ by using~(\ref{equ:quantization})
\begin{equation}
    \ak = \frac{g_k^*(\tau)\hat{\zeta}_{\vk}(\tau)+if_k^*(\tau)\hat{\pi}_{\vk}(\tau)}{f_k(\tau)g_k^*(\tau)+f_k^*(\tau)g_k(\tau)}\hs,
\end{equation}
also, recall the following formulas
\begin{equation}
\begin{aligned}
    \hat U(\tau,\tau_0) \hat{A}(\tau) \hat U^{\dagger}(\tau,\tau_0) &= \hat{A}^{\mathrm{S}}(\tau)= \hat{A}(\tau_0)\hs,\\
    \sandwitch{\zeta'}{\hat{\zeta}_{\vk}(\tau_0)}{\zeta}&=\bar{\zeta}_{k}\delta(\zeta'-\zeta)\hs,\\
    \sandwitch{\zeta'}{\hat{\pi}_{\vk}(\tau_0)}{\zeta}&=-i\delta(\zeta'-\zeta)\frac{\partial}{\partial \bar{\zeta}_{k}^*}\hs,\\
\end{aligned}
\end{equation}
where we use the superscript $\mathrm{S}$ to represent Schr\"{o}dinger picture. Subsequently, we can rewrite the differential equation for the vacuum state in the field basis 
\begin{equation}
    \forall \vk: \bar{\zeta}_{\vk}\Psi_{G}[\zeta]+\frac{f^*_k(\tau)}{g^*_k(\tau)}\frac{\partial}{\partial\bar{\zeta}_{-\vk}}\Psi_{G}[\zeta]=0\hspace{2pt},
\end{equation}
where $\Psi_{G}[\zeta]=\braket{\zeta|\Omega(\tau)}$. By solving this equation we get the vacuum state for a fixed mode
\begin{equation}
    \Psi_{G}[\zeta_{\vk},\zeta_{-\vk}] = N_{\zeta_{\vk}}(\tau)\exp[-A_{\zeta}(k,\tau)|\bar{\zeta}_{\vk}|^2]\hspace{2pt},
\label{equ:initial state}
\end{equation}
where $N_{\zeta_{\vk}}(\tau)$ is the normalization parameter, and $A_{\zeta}(k,\tau)$ is defined as 
\begin{equation}
A_{\zeta}(k,\tau) \equiv \frac{g^*_k(\tau)}{f^*_k(\tau)} =2k^3   \frac{\epsilon M_{pl}^2}{H^2}  \frac{1-\frac{i}{k\tau}}{1+k^2\tau^2}\hspace{2pt}.
\label{equ:dynamics term}
\end{equation}
To obtain the complete wave functional, we have relation\footnote{Since $\vk$ and $-\vk$ modes share the same solution $\Psi_G[\zeta_{\vk}]$, the sum only includes positive part.}
\begin{equation}
    \Psi_{G}[\zeta]=\prod_{\mathbf{k}\in\mathbb{R}^{3+}}\Psi_{G}[\zeta_{\mathbf{k}},\zeta_{-\vk}]=\left(\prod_{\mathbf{k}\in \mathbb{R}^{3+}}N_{\zeta_{\vk}}(\tau)\right)\exp\left[-\sum_{\mathbf{k}\in \mathbb{R}^{3+}}A_{\zeta}(k,\tau)|\bar{\zeta}_{\vk}|^2\right]\hspace{2pt}.
\end{equation}
We can redefine the normalization parameter $N_{\zeta}(\tau)\equiv\left(\prod_{\mathbf{k}\in \mathbb{R}^{3+}}N_{\zeta_{\vk}}(\tau)\right)$ and convert the summation into an integration via $\frac{\sum_{\mathbf{k}}}{V}=\int\frac{d^3\mathbf{k}}{(2\pi)^3}$. Subsequently, the complete wave functional is
\begin{equation}
  \Psi_{G}[\zeta] = N_{\zeta}(\tau)\exp\left[-\frac{1}{2}\int\frac{d^3k}{(2\pi)^3}A_{\zeta}(k,\tau)|\zeta_{\vk}|^2\right]\hspace{2pt},
\end{equation}
where $\zeta_{\mathbf{k}}=\sqrt{V}\bar{\zeta}_{\mathbf{k}}$, $V=(2\pi)^3\delta(\mathbf{0})$ and $\frac{1}{2}$ comes from the over counting since the integral is over the whole $\mathbb{R}^{3}$.

\subsection{Interaction and decoherence}
Then we consider quantum decoherence, which is the loss of quantum coherence and is captured by the loss of quantum information. The core of quantum decoherence is the entanglement between the environment and the system. Since we cannot obtain the full information of the environment, we need to trace it out and this process can be understood as a measurement. Through it, we can reconstruct the quantum state of the system, which is called the reduced density matrix.

In an ideal measurement, only the diagonal elements remain the same, and all the information contained in the off-diagonal elements is lost. Therefore, the reduced density matrix represents a mixed state, and one cannot distinguish such a state from a classical stochastic distribution. Since the main (and the only in an ideal case) difference between the original state and the reduced density matrix is the information contained in the off-diagonal elements, we need an observable $\hat{\mathcal{O}}$ that is not diagonal to distinguish them. In the cosmology context, the n-point correlation function $\braket{\zeta^n}$ is trivial and does not carry any off-diagonal information. Therefore we introduce the Bell operator which contains off-diagonal elements to test the decoherence. 

In this paper, we consider decoherence from gravitational nonlinearity~\cite{Nelson:2016kjm} which can be understood as the gravitational backreaction of the fluctuation of the inflaton. It provides us a lower bound of the decoherence effect since even without any additional fields and interactions, we can still have gravity. Expand the action of gravity and a scalar field up to cubic terms and the leading bulk term\footnote{(\ref{equ:bulk_cubic_term}) is also the leading contribution compared with self-interaction which is suppressed by slow-roll condition. One can refer \cite{Sou:2022nsd} for a concrete calculation.} for decoherence is \cite{Maldacena:2002vr,Nelson:2016kjm} 
\begin{equation}
\mathcal{L}^{\mathrm{bulk}}_{\mathrm{int}}=\epsilon(\epsilon+\eta) a(\tau)^2 M_{pl}^2\zeta\left(\partial_i \zeta\right)^2+...\hspace{2pt}, \label{equ:bulk_cubic_term}
\end{equation}
while the leading boundary term for scalar on the superhorizon scale is \cite{Maldacena:2002vr,Sou:2022nsd}
\begin{equation}  \mathcal{L}^{\mathrm{bd}}_{\mathrm{int}}=\partial_{\tau}\left(-2 a^3 H M_{pl}^2 e^{3 \zeta}\right)+...\hspace{2pt}.
\label{equ:boundary term}
\end{equation}
In the following, we review the formalism in \cite{Nelson:2016kjm} to calculate the decoherence rate from the above interaction. We treat one particular mode $\zeta_S\equiv\xi$ as a system\footnote{To be rigorous, one can consider the system to be the observable modes with comoving momentum $q_{\rm min}<q<q_{\rm max}, $, but the dependence on $q_{\rm min}$ and $q_{\rm max}$ only affects sub-dominated (in scale factor $a(\tau)$) terms in decoherence rate, as shown in \cite{Burgess:2022nwua,Ning:2023ybc}.}, and other wavelength modes $\zeta_E\equiv\mathcal{E}$ as an environment. In the bulk case, the dominant term is
\begin{equation}
\zeta\left(\partial_i\zeta\right)^2\rightarrow\xi(\partial_i \mathcal{E})^2\hspace{2pt},
\end{equation}
since compared to another type of interaction between the system and environment $\xi\xi\mathcal{E}$, the type $\mathcal{E}_{\bf k}\mathcal{E}_{{\bf k}'}\xi_{\bf q}$ allows much more modes with the conservation of momentum ${\bf k}+{\bf k}'+{\bf q}={\bf 0}$, and such a fact also applies for the cubic boundary interactions.

The above bulk interaction in the momentum space has the form 
\begin{equation}
    H_{\mathrm{int}}(\tau)\supset \int_{\mathbf{k},\mathbf{k'},\mathbf{q}}\tilde{\mathcal{H}}_{\mathbf{k}, \mathbf{k}^{\prime}, \mathbf{q}}^{(\mathrm{int})}(\tau)\mathcal{E}_{\mathbf{k}}\mathcal{E}_{\mathbf{k}'}\xi_{\mathbf{q}}\hspace{2pt},
\label{equ:bulk interaction in momentum space}
\end{equation}
where$\int_{\mathbf{k}, \mathbf{k}^{\prime}, \mathbf{q}} \equiv \int \frac{d^3 \mathbf{k}}{(2 \pi)^3} \frac{d^3 \mathbf{k}^{\prime}}{(2 \pi)^3} \frac{d^3 \mathbf{q}}{(2 \pi)^3}(2 \pi)^3 \delta^3\left(\mathbf{k}+\mathbf{k}^{\prime}+\mathbf{q}\right)$, $\tilde{\mathcal{H}}_{\mathbf{k}, \mathbf{k}^{\prime}, \mathbf{q}}^{(\mathrm{int})}$ is the coefficient of the interaction in momentum space. Then we consider the wave functional evolution of the vacuum state $\Psi[\zeta]$ for generic cubic interaction.  We can make the following ansatz
\begin{equation}   \Psi[\xi,\mathcal{E}]=\Psi_G[\xi]\Psi_G[\mathcal{E}]\exp{\left(\int_{\mathbf{k},\mathbf{k'},\mathbf{q}}\mathcal{F}_{\mathbf{k},\mathbf{k'},\mathbf{q}}\mathcal{E}_{\mathbf{k}}\mathcal{E}_{\mathbf{k'}}\xi_{\mathbf{q}}\right)}\hspace{2pt},
\label{equ:ansatz}
\end{equation}
where $\Psi_{G}[\xi]$ and $\Psi_G[\mathcal{E}]$ are the wave functional~(\ref{equ:initial state}) corresponding to the free action. Through the Schr\"{o}dinger equation we have
\begin{equation}
\mathcal{F}_{\mathbf{k}, \mathbf{k}^{\prime}, \mathbf{q}}(\tau)=i \int_{\tau_0}^\tau \frac{d \tau^{\prime}}{H \tau^{\prime}} \tilde{\mathcal{H}}_{\mathbf{k}, \mathbf{k}^{\prime}, \mathbf{q}}^{(\mathrm{int})}\left(\tau^{\prime}\right) \exp \left[i \int_{\tau^{\prime}}^\tau d \tau^{\prime \prime} \alpha_{k, k^{\prime}, q}\left(\tau^{\prime \prime}\right)\right]\hspace{2pt},
\end{equation}
and $\alpha_{k, k^{\prime}, q}\left(\tau^{\prime \prime}\right)$ is defined as
\begin{equation}
\alpha_{k, k^{\prime}, q}(\tau)=\frac{1}{H \tau}\left[f_{\mathcal{E}}(k, \tau) A_{\mathcal{E}}(k, \tau)+\left(k \rightarrow k^{\prime}\right)\right]+\frac{1}{H \tau} f_\xi(q, \tau) A_\xi(q, \tau)\hspace{2pt},
\end{equation}
where $f_{\mathcal{E}}(k,\tau)$ and $f_{\xi}(k,\tau)$ are the coefficient of the kinetic term in momentum space and $A_{\mathcal{E}}(k,\tau)$ and $A_{\xi}(k,\tau)$ are just the coefficient defined in~(\ref{equ:initial state}). In general, the system and the environment can have different dynamics, so we use subscripts to distinguish them. However, in our case, since they are just different modes of the scalar perturbation $\zeta$, we have $ f_{\mathcal{E}}(k,\tau)=f_{\xi}(k,\tau)=-\frac{H^3 \tau^3}{2 \epsilon M_{pl}^2}$ and $A_{\mathcal{E}}(k,\tau)=A_{\xi}(k,\tau)$ given by~(\ref{equ:dynamics term}).

For the boundary interaction case, there is a more convenient way to calculate the modified wave functional which is still a product of a Gaussian part and an extra phase, and the result is given by \cite{Sou:2022nsd}, and see also \cite{Ning:2023ybc} for the derivation based on the WKB approximation of the Wheeler-DeWitt equation. In this case, at the cubic order, we have Hamiltonian:
\begin{equation}
    H_{\rm bd}(\zeta,\tau) = \partial_{\tau} K(\zeta,\tau)\hspace{2pt},
\end{equation}
and the corresponding wave functional is 
\begin{equation}   \Psi[\zeta]=\Psi_G[\zeta]\exp{\left[-iK(\zeta,\tau)\right]}\hspace{2pt}.
\label{equ: bd wavenfunctional}
\end{equation}
Subsequently, expanding the phase in momentum space to cubic term, we will go back to~(\ref{equ:ansatz}).

The reduced density matrix of system $\xi$ is obtained by integration of the environment $\mathcal{E}$
\begin{equation}
    \rho_R[\xi,\tilde{\xi}]=\Psi_G[\xi]\Psi^*_G[\tilde{\xi}]\braket{\exp{\left(\int_{\mathbf{k},\mathbf{k'},\mathbf{q}}\mathcal{E}_{\mathbf{k}}\mathcal{E}_{\mathbf{k'}}\left(\xi_{\mathbf{q}}\mathcal{F}_{\mathbf{k},\mathbf{k'},\mathbf{q}}+\tilde{\xi}_{\mathbf{q}}\mathcal{F}^*_{\mathbf{k},\mathbf{k'},\mathbf{q}}\right)\right)}}_{\mathcal{E}}\hspace{2pt},
\end{equation}
where
\begin{equation}
\langle\cdots\rangle_{\mathcal{E}} \equiv \int D \mathcal{E}\left|\Psi_G^{(\mathcal{E})}\right|^2(\cdots)\hspace{2pt}, \label{eq:tracing_envir}
\end{equation}
Make use of the following formula
\begin{equation}
\left\langle e^X\right\rangle=\exp \left[\frac{1}{2}\left\langle X^2\right\rangle_c+\frac{1}{4 !}\left\langle X^4\right\rangle_c+\ldots\right]\hspace{2pt},
\label{eq:moments expansion}
\end{equation}
where the subscript $_{c}$ means the connected part of the correlation function. We keep only the leading order and neglect the real part of $\mathcal{F}$ which should be very small to keep the density matrix close to Gaussian~\cite{Nelson:2016kjm}. Then we can parameterize the reduced density matrix in the form
\begin{equation}  \rho_R[\xi_{\mathbf{q}},\tilde{\xi}_{\mathbf{q}}]=\Psi_G[\xi_{\mathbf{q}}]\Psi^*_G[\tilde{\xi}_{\mathbf{q}}]\exp \left(-\tilde{\Gamma}|\Delta \bar{\xi}_{\mathbf{q}}|^2\right)\hspace{2pt},
\label{equ:reduce density matrix}
\end{equation}
where $\left|\Delta\bar{\xi}_{\mathbf{q}}\right|=\left|\bar{\xi}_{\mathbf{q}}-\bar{\tilde{\xi}}_{\mathbf{q}}\right|$ the same ansatz is also considered phenomenologically in \cite{Kiefer:1999sj,Kiefer:2006jea} and appears in the quantum master equation formalism of decoherence \cite{Hollowood:2017bil}. The decoherence factor $\tilde{\Gamma}$ associates with the decoherence rate $\Gamma$ which is defined as
\begin{equation}
    \frac{4\pi^2\Delta_{\zeta}^2}{q^3}\tilde\Gamma \equiv \Gamma =\frac{4\pi^2\Delta_{\zeta}^2}{q^3}\int_{\mathbf{k}+\mathbf{k}^{\prime}=-\mathbf{q}} P_{\mathcal{E}}\left(k, \tau\right) P_{\mathcal{E}}\left(k^{\prime}, \tau\right)\left(\operatorname{Im} \mathcal{F}_{\mathbf{k}, \mathbf{k}^{\prime}, \mathbf{q}}(\tau)\right)^2\hspace{2pt},
\end{equation}
where $\Delta_{\zeta}^2 = \frac{H^2}{8\pi^2\epsilon M_{pl}^2}$ is the dimensionless power spectrum and $P_{\mathcal{E}}\left(k, \tau\right)$ is the power spectrum, and from the definition it is 
\begin{equation}
    P_{\mathcal{E}}\left(k, \tau\right) = \frac{1}{2\operatorname{Re}A_{\mathcal{E}}(k,\tau)}\hspace{2pt}.
\end{equation}
Here we list the value of $\Gamma$ for the bulk term decoherence (\ref{equ:bulk_cubic_term}) and boundary term decoherence (\ref{equ:boundary term}) respectively~\cite{Nelson:2016kjm,Sou:2022nsd}
\begin{align}
\Gamma_{\mathrm {bulk}}&=\frac{4\pi^2\Delta_{\zeta}^2}{8 \pi}\left\{\left(\frac{\epsilon+\eta}{12}\right)^2 \left(\frac{a H}{q}\right)^3+\frac{(\epsilon+\eta)^2}{9 \pi}\left(\frac{a H}{q}\right)^2\left[\Delta N-\frac{19}{48}\right]\right\}\hspace{2pt},\\
\Gamma_{\mathrm{bd}}&\approx \frac{729 \Delta_{\zeta}^2 }{16 \epsilon^2}\left[4(\Delta N-1)\left(\frac{a H}{q}\right)^6-4 \Delta N\left(\frac{a H}{q}\right)^4+\frac{5 \pi}{2}\left(\frac{a H}{q}\right)^3+(4 \Delta N-7)\left(\frac{a H}{q}\right)^2\right]  \hspace{2pt}. 
\end{align}
where $\Delta N = \log\left(\frac{q}{k_{\mathrm{min}}}\right)$, the e-folds from the beginning of the inflation to the horizon crossing of the mode $\mathbf{q}$, is a cutoff to resolve the IR divergence from the 1-loop contribution of the massless comoving curvature perturbation. The decoherence rate from the boundary term listed here is only a late-time limit approximation, and the full function can be found in~\cite{Sou:2022nsd}. 

\subsection{Bell inequalities}
Let us recall the simplest setup of Bell inequalities with a pair of spin-1/2 particles. Consider two sets of operators $\hat{A}, \hat{A}'$ and $\hat{B}, \hat{B}'$. They act in different Hilbert spaces $\mathcal{H}_A$ and $\mathcal{H}_B$ (i.e. particle A and particle B). These operators correspond to measuring the spin along various axes, which can be labeled by a unit vector $\mathbf{n}$. Therefore, those operators can be written as $\hat{A}= n_i\cdot\sigma_A^i$,  and $\hat{A}', \hat{B}, \hat{B}'$ are defined similarly, where $\sigma^i$ are Pauli matrices. It is easy to see that in the spin 1/2 system, all those operators have eigenvalues $\pm1$.

Then as Clauser, Horne, Shimony and Holt (CHSH) introduced~\cite{Clauser:1969ny}, we can construct the following Bell operator
\begin{equation}
    \hat{\mathcal{B}}=\hat{A} \otimes \hat{B}+\hat{A}^{\prime} \otimes \hat{B}+\hat{A} \otimes \hat{B}^{\prime}-\hat{A}^{\prime} \otimes \hat{B}^{\prime}\hspace{2pt}.
\end{equation}
The local hidden variable theories give the following results.
\begin{equation}
|\langle{\hat{\mathcal{B}}}\rangle|\leq 2\hspace{2pt},  
\end{equation}
which can be easily seen if we rewrite the expression as:
    $\hat{\mathcal{B}}=\hat{A} \otimes\left(\hat{B}+\hat{B}^{\prime}\right)+ \hat{A}^{\prime} \otimes\left(\hat{B}-\hat{B}^{\prime}\right)$
and noticed that we either have $\hat{B}=\hat{B}'$ or $\hat{B}=-\hat{B}'$. However, in quantum mechanics, if $\hat{B}$ and $\hat{B}'$ are noncommuting, we cannot determine their values simultaneously. Therefore if we have non-commuting $\hat{A}, \hat{A}'$ and $\hat{B}, \hat{B}'$, we may have a bigger value of the expectation of $\hat{\mathcal{B}}$, namely the violation of Bell inequalities. To make it clear, we can check its square.
\begin{equation}
    \hat{\mathcal{B}}^2 = 4I-\left[\hat{A}, \hat{A}'\right]\left[\hat{B}, \hat{B}'\right]\hspace{2pt},
\end{equation}
where $I$ is the unit matrix and we used $\hat{A}^2=1, \hat{A}'^2=1$, etc. As mentioned above, the non-zero commutator term can make $\langle\hat{\mathcal{B}}^2\rangle$ bigger than four. One can prove $\left|\left[\hat{A}, \hat{A}'\right]\right|\leq 2$ and $\left|\left[\hat{B}, \hat{B}'\right]\right|\leq 2$ . Therefore in quantum mechanics, the maximal violation of Bell Inequalities is $2\sqrt{2}$ which is called the Cirel'son bound \cite{Cirelson:1980ry}.

\subsection{Bell operators in field theory and cosmology}
To consider the Bell inequality in cosmology, we also need to construct a Bell operator. However, now the Hilbert space we are facing is more complicated. In quantum field theory, we can expand the field by its modes, and each mode can be seen as a harmonic oscillator. If the Hilbert space corresponding to a harmonic oscillator is $\mathcal{H}_{\mathrm{HO}}$, then the total Hilbert space of the comoving curvature perturbation is $\Pi_{\mathbf{k}\in\mathbb{R}^{3+}}\mathcal{H}_{\mathrm{HO},\mathbf{k}}\otimes\mathcal{H}_{\mathrm{HO},\mathbf{-k}}$. Fortunately, there exist several constructions called pseudo-spin operators which have the same algebraic structure with Pauli matrices for such system, like the Banaszek-Wodkiewicz (BW) pseudo-spin operators~\cite{banaszek:1999,chen:2002} and the Gour-Khanna-Mann-Revzen (GKMR) pseudo-spin
operators~\cite{revzen:2005,gour:2004}.

In this paper, we use the GKMR operators to construct Bell operators for its convenience when calculating its matrix elements in the representation of field operator eigenstates. Let us begin with creation and annihilation operators $\admk,\ak$ for a given mode $\mathbf{k}$, and we can introduce position operator and momentum operator
\begin{equation}
\begin{aligned}
    \hat{x}_{\mathbf{k}}=\frac{1}{\sqrt{2k}}(\ak+\adk)\hs,\\
    \hat{p}_{\mathbf{k}}=-i\sqrt{\frac{k}{2}}(\ak-\adk)\hs,
\label{equ: single mode variables}
\end{aligned}
\end{equation} 
as we do in the harmonic oscillator model\footnote{Here we abuse the notation of the creation and annihilation operators. Here they are defined on the same time slice with $\hat{x}$ like $\tau$ while in (\ref{equ:quantization}) they are defined on a fixed time slice $\tau_0$. Later when we have to treat them at the same time we will label them $\ak(\tau)$ and $\ak(\tau_0)$ respectively.}. Then we use $\left|\mathcal{E}_{\mathbf{k}}\right\rangle$ and $\left|{\mathcal{O}_{\mathbf{k}}}\right\rangle$
\begin{equation}
    \begin{aligned}
    \left|\mathcal{E}_{\mathbf{k}}\right\rangle & =\frac{1}{\sqrt{2}}\left(\left|x_{\mathbf{k}}\right\rangle+\left|-x_{\mathbf{k}}\right\rangle\right) \hspace{2pt},\\
    \left|\mathcal{O}_{\mathbf{k}}\right\rangle & =\frac{1}{\sqrt{2}}\left(\left|x_{\mathbf{k}}\right\rangle-\left|-x_{\mathbf{k}}\right\rangle\right)\hspace{2pt},
\end{aligned}
\end{equation}
to define the GKMR pseudo-spin operators
\begin{align}
\hat{S}_{x}(\mathbf{k})&=\int_{0}^{+\infty} d x_{\mathbf{k}}\left(\left|\mathcal{E}_{\mathbf{k}}\right\rangle\left\langle\mathcal{O}_{\mathbf{k}}|+| \mathcal{O}_{\mathbf{k}}\right\rangle\left\langle\mathcal{E}_{\mathbf{k}}\right|\right)\label{Sx}\hspace{2pt}, \\
\hat{S}_{y}(\mathbf{k})&=i \int_{0}^{+\infty} d x_{\mathbf{k}}\left(\left|\mathcal{O}_{\mathbf{k}}\right\rangle\left\langle\mathcal{E}_{\mathbf{k}}|-| \mathcal{E}_{\mathbf{k}}\right\rangle\left\langle\mathcal{O}_{\mathbf{k}}\right|\right)\hspace{2pt}, \\
\hat{S}_{z}(\mathbf{k})&=-\int_{0}^{+\infty} d x_{\mathbf{k}}\left(\left|\mathcal{E}_{\mathbf{k}}\right\rangle\left\langle\mathcal{E}_{\mathbf{k}}|-| \mathcal{O}_{\mathbf{k}}\right\rangle\left\langle\mathcal{O}_{\mathbf{k}}\right|\right)\label{Sz}\hspace{2pt},    
\end{align}
where one can verify that these operators satisfy the $SU(2)$ commutation relation that $\left[\hat{S}_i, \hat{S}_j\right]=2i\epsilon_{ijk}\hat{S}_k~(i,j,k=1,2,3)$, and have eigenvalues $\pm 1$. Notice that the field operators mix modes $\mathbf{k}$ and $-\mathbf{k}$ while $\hat{x}_{\mathbf{k}}$ and $\hat{p}_{\mathbf{k}}$ act only on $\mathcal{E}_{\mathbf{k}}$. 

Now we can naturally define the Bell operator for the continuous variable systems as we do in the spin-1/2 system. Since we see that the quantum state of the fluctuation is a two-mode squeezed state on $\mathcal{E}_{\mathbf{k}}\otimes\mathcal{E}_{\mathbf{-k}}$, we let $\hat{A},\hat{A}'$ acting on $\mathcal{E}_{\mathbf{k}}$ and $\hat{B}, \hat{B}'$ acting on $\mathcal{E}_{-\mathbf{k}}$. Then we only need to specify two sets of unit vectors $\mathbf{n}, \mathbf{n}'$ and $\mathbf{m}, \mathbf{m'}$. For two vectors with the same origin, we can always put them in the $(x,z)$
plane. Therefore, we have $\mathbf{n}=(\sin{\theta_{n}},0,\cos{\theta_{n}})$, $\hat{A}=n^i\hat{S}_i(\mathbf{k})=\sin{\theta_n}\hat{S}_x(\mathbf{k})+\cos{\theta_n}\hat{S}_z(\mathbf{k}) $, and $\mathbf{n}', \mathbf{m}, \mathbf{m'}$  as well as $\hat{A}',\hat{B},\hat{B}'$ can be expressed similarly. To completely determine the Bell operator we still need to specify these four angles $\theta_n,\theta_{n'},\theta_{m},\theta_{m'}$,  which is an optimization problem. Here, we choose a configuration that $\theta_n=0, \theta_{n'}=\pi/2, \theta_m=-\theta_{m'}=\arctan\left[\left\langle \hat{S}_x(\mathbf{k})\hat{S}_x(-\mathbf{k})\right\rangle/\left\langle \hat{S}_z(\mathbf{k})\hat{S}_z(-\mathbf{k})\right\rangle\right] $. Here we used $\left\langle \hat{S}_x(\mathbf{k})\hat{S}_z(\mathbf{-k})\right\rangle=0$  and $\left\langle \hat{S}_z(\mathbf{k})\hat{S}_x(\mathbf{-k})\right\rangle=0$ which can be obtained through straightforward calculation and we will calculate later.
So, the expectation value of the Bell operator is given by
\begin{equation}
\left\langle\hat{\mathcal{B}}\right\rangle=2\sqrt{\left\langle \hat{S}_x(\mathbf{k})\hat{S}_x(-\mathbf{k})\right\rangle^2+\left\langle \hat{S}_z(\mathbf{k})\hat{S}_z(-\mathbf{k})\right\rangle^2}\hspace{2pt}.
\label{equ:Bell operator}
\end{equation}
\section{Effects of quadratic boundary terms}
\label{sec: effect of quadratic}
In this section, we first discuss the change of the wave functional of the fluctuation when considering a quadratic boundary which can be understood as a canonical transformation. A derivation and an example are given to explain why some quantities like particle number change under the canonical transformation.

\subsection{Quadratic boundary terms and canonical transformations}
First, from~(\ref{equ:boundary term}) and~(\ref{equ: bd wavenfunctional})\footnote{For the third order interaction, the Hamiltonian is just flipping the sign of the third order interaction. One can refer \cite{Wang:2013zva} for detailed derivation. For a more systematic derivation at higher order, we may apply the WKB approximation to the Wheeler-DeWitt equation to solve the corresponding Hamilton-Jacobi equation \cite{Ning:2023ybc}.}, the full wave functional when considering the boundary term is given by:
\begin{equation}
  \Psi^{(b)}[\zeta] = N_{\zeta}(\tau)\exp\left(-\frac{1}{2}\int\frac{d^3k}{(2\pi)^3}A_{\zeta}(k,\tau)|\zeta_{\vk}|^2\right)\exp\left(-2ia^3HM_{pl}^2\int d^3x e^{3\zeta}\right)\hspace{2pt},
  \label{equ:new vacuum}
\end{equation}
where superscript $(b)$ refers to the objects (quantum states, observable, etc.) after adding the boundary term, in other words, after applying the canonical transformation. 

Expand the boundary-related term to quadratic order and we find that it contributes to the imaginary part of $A_{\zeta}$:
\begin{align}
        \Psi^{(b)}[\zeta] &= N_{\zeta}(\tau)\exp\left(-\frac{1}{2}\int\frac{d^3k}{(2\pi)^3}A^{(b)}_{\zeta}(k,\tau)|\zeta_{\vk}|^2\right)\hspace{2pt},\\
        A_{\zeta}^{(b)}&\equiv A_{\zeta}+18ia^3 H M_{pl}^2\label{equ: dynamic term with bd}\hs.
\end{align}
This is in fact a canonical transformation
\begin{equation}
\begin{aligned}
    \hat{\mathcal{S}}\hat{\zeta}\hat{\mathcal{S}}^{-1} &\rightarrow \hat{\zeta}\hs,\\
    \hat{\mathcal{S}}\hat{\pi}\hat{\mathcal{S}}^{-1} &\rightarrow \hat{\pi} - 18a^3HM_{pl}^2\hat{\zeta}\hs,
    \label{equ: canonical transformation}
\end{aligned}
\end{equation}
which is generated by quadratic boundary action
\begin{equation}
\begin{aligned}
    \mathcal{L}_{\mathrm{bd}}&=\partial_\tau\left(-9a^3HM_{pl}^2\zeta^2\right)\hs,\\
    \pi^{(b)}&=\frac{\partial\mathcal{L}_0}{\partial\zeta'}+\frac{\partial\mathcal{L}_{\mathrm{bd}}}{\partial\zeta'}=\pi - 18a^3HM_{pl}^2\zeta\hs. \label{equ:quadratic_bd_term}
\end{aligned}
\end{equation}
In field basis, for each mode $k$ the canonical transformation can be expressed as\footnote{This is the last time we clearly distinguish $\zeta_\vk$ and $\bar{\zeta}_{\vk}$. We will neglect the $\bar{~}$ for simplicity, which is equivalent to absorbing the factor $(2\pi)^3\delta(0)$, in the rest of the paper.}
\begin{equation}
    \mathcal{S}|_{k} = \exp{(-18ia^3HM_{pl}^2 |\bar\zeta_{\vk}|^2)}\hs,
\end{equation}
one can take it into~(\ref{equ: canonical transformation}) to verify and therefore
\begin{equation}
    \mathcal{S}=\exp\left[-\frac{1}{2}\int\frac{d^3k}{(2\pi)^3}18ia^3HM_{pl}^2 |\zeta_{\vk}|^2\right]\hs,
\end{equation}
where $\frac{1}{2}$ comes from the same reason as wave functional and this is exactly the extra phase factor in~(\ref{equ:new vacuum}).

Under the canonical transformation, there should be no change to physics. Therefore, one may expect any quantity $\hat{\mathcal{O}}$ is invariant since
\begin{equation}
    \sandwitch{\Psi}{\hat{\mathcal{O}}}{\Psi} = \sandwitch{\Psi^{(b)}}{\hat{\mathcal{S}}\hat{\mathcal{O}}\hat{\mathcal{S}}^{-1}}{\Psi^{(b)}}\hs,
\end{equation}
as long as $\hat{\mathcal{S}}\hat{\mathcal{O}}\hat{\mathcal{S}}^{-1}=\hat{\mathcal{O}}^{(b)}$ holds. However, this is not always true, due to the mismatching of the canonical transformation on states and operators. In the following, we discuss explicitly how the difference arises. 
\begin{figure}
    \centering
\begin{tikzcd}
&
\{\hat{\zeta},\hat{\pi}\}(\tau)
\ar{dl}[sloped]{\mathcal{D}}
\ar{rr}[near end]{\hat{\mathcal{S}}_{\tau}}
& & \{\hat{\zeta}^{(b)},\hat{\pi}^{(b)}\}(\tau)
\ar{dl}[swap, sloped, near start]{\mathcal{D}^{(b)}}
\\
\{\hat{a}_{\mathbf{k}},\hat{a}^{\dagger}_{-\mathbf{k}}\}(\tau)
\ar{rr}[near end]{\hat{\tilde{\mathcal{S}}}_{\tau}}
& & \{\hat{b}_{\mathbf{k}},\hat{b}^{\dagger}_{-\mathbf{k}}\}(\tau)
\\
&
\{\hat{\zeta},\hat{\pi}\}(\tau_0)
\ar[]{uu}[near start, crossing over]{\hat{U}(\tau, \tau_0)}
\ar{rr}[near end, crossing over]{\hat{\mathcal{S}}_{\tau_0}}
\ar[sloped]{dl}{\mathcal{D}}
& & \{\hat{\zeta}^{(b)},\hat{\pi}^{(b)}\}(\tau_0)
\ar{dl}[sloped,swap]{\mathcal{D}^{(b)}}
\ar{uu}[near start,swap]{\hat{U}^{(b)}(\tau, \tau_0)}
\\
\{\hat{a}_{\mathbf{k}},\hat{a}^{\dagger}_{-\mathbf{k}}\}(\tau_0)
\ar{uu}[near start]{\hat{U}(\tau, \tau_0)}
\ar{rr}[near end]{\hat{\tilde{\mathcal{S}}}_{\tau_0}}
& & \{\hat{b}_{\mathbf{k}},\hat{b}^{\dagger}_{-\mathbf{k}}\}(\tau_0)
\ar{uu}[near start,swap]{\hat{U}^{(b)}(\tau, \tau_0)}
\end{tikzcd}
    \caption{Canonical transformation map. $\hat{U}$ is time evolution, $\hat{\mathcal{S}}$ is canonical transformation, and $\mathcal{D}$ maps canonical variables to creation and annihilation operators. The left and right surfaces represent different canonical variables with time increasing from bottom to top. The canonical transformation induced by the quadratic boundary term is $\hat{\mathcal{S}}_{\tau}$, however in Schr\"{o}dinger picture, operators do not change with time and fix in the time $\tau_0$ when the vacuum is defined in Heisenberg picture. Therefore, in general the canonical transformation of observables $\hat{\mathcal{S}}_{\tau_0}$ or $\hat{\tilde{\mathcal{S}}}_{\tau_0}$ is not consistent with the canonical transformation of quantum states $\hat{\mathcal{S}}_{\tau}$.}
    \label{fig:canonical transformation map}
\end{figure}

It is convenient to use Fig.~\ref{fig:canonical transformation map}, which shows all the operators and their relationships we are going to use. In this cube, there are 3 types of transformation, time evolution $\hat{U}$, canonical transformation $\hat{\mathcal{S}}$ and $\mathcal{D}$ which maps canonical variables to creation and annihilation operators. First, notice two side surfaces of the cube. In general, the choice of the $\mathcal{D}$ is ambiguous, the only requirement is that the transformation should be compatible with the symplectic structure (i.e. the canonical commutation relation between creation and annihilation operators). 

However, conventionally the choice of $\mathcal{D}$ is contained in the following quantization process
\begin{equation}
\begin{aligned}
    \hat{\zeta}_{\vk}(\tau) &= f_k(\tau)\ak(\tau_0)+ f_k^*(\tau)\admk(\tau_0)\hs,\\
    \hat{\pi}_{\vk}(\tau) &= -i\left[g_k(\tau)\ak(\tau_0)-g_k^*(\tau)\admk(\tau_0)\right]\hs,
\end{aligned}
\end{equation}
which is just~(\ref{equ:quantization}), and $f_k$, $g_k$ follow the classical solution of canonical variable and corresponding conjugate momentum. $\mathcal{D}$ can be expressed by $f_k(\tau_0)$ and $g_k(\tau_0)$. Therefore, for different choices of canonical variables the choice of $\mathcal{D}$ would be different. So in the figure, we label them $\mathcal{D}$ and $\mathcal{D}^{(b)}$ respectively. Second, notice the front and back surfaces of the cube. If the canonical transformation depends on time, $\hat{U}\neq\hat{U}^{(b)}$. Finally, the top and bottom surfaces show that the canonical transformation for canonical variables $\hat{\mathcal{S}}$ is different from the transformation $\hat{\tilde{\mathcal{S}}}$ between the creation and annihilation operators if $\mathcal{D}\neq{\mathcal{D}^{(b)}}$.

Now, back to cosmological perturbation case, $\mathcal{S}_{\tau}=\exp\left[-\frac{1}{2}\int\frac{d^3k}{(2\pi)^3}18ia(\tau)^3HM_{pl}^2 |\zeta_{\vk}(\tau_0)|^2\right]$ (to make it clear, in this part we will label all the time dependence of the quantities) which is time dependent. The initial time $\tau_0$ is usually $\tau_0\rightarrow-\infty$ which is the subhorizon limit and the spacetime should look like Minkowski. Therefore, we have $\mathcal{D}=\mathcal{D}^{(b)}$ and we can also get this result from an observation that, $\hat{\mathcal{S}}_{\tau_0}=\hat{\tilde{\mathcal{S}}}_{\tau_0}=\hat{I}$ an identity operator, since $a(\tau_0)=0$. Recall in the Schr\"{o}dinger picture, that operators do not evolve with time, and from Fig.~\ref{fig:canonical transformation map} we know
\begin{equation}
\begin{aligned}
    \hat{\mathcal{S}}_{\tau}\mathcal{O}_{\zeta}\left(\hat{\zeta}_{\vk}(\tau_0),\hat{\pi}_{\vk}(\tau_0)\right)\hat{\mathcal{S}}^{-1}_{\tau}\neq \mathcal{O}_{\zeta}\left(\hat{\zeta}_{\vk}^{(b)}(\tau_0),\hat{\pi}_{\vk}^{(b)}(\tau_0)\right)\hs,\\
    \hat{\mathcal{S}}_{\tau}\mathcal{O}_{a}\left(\ak(\tau_0),\admk(\tau_0)\right)\hat{\mathcal{S}}^{-1}_{\tau}\neq \mathcal{O}_{a}\left(\bk(\tau_0),\bdmk(\tau_0)\right)\hs,
\end{aligned}
\end{equation}
where $\mathcal{O}_{\zeta}$ is a combination of operators $\hat{\zeta}_{\vk}(\tau_0),\hat{\pi}_{\vk}(\tau_0)$
 and $\mathcal{O}_{a}$ is a combination of operators $\ak(\tau_0),\admk(\tau_0)$. There are two remarks.
 First, the combination could be either a linear or nonlinear combination of operators, and for the later case, the order should be assigned. The coefficients of the combination can be a function of time but the coefficients do not change under time evolution or canonical transformation. Second, the equality recovers for trivial $\mathcal{O}$ which means for any time $\tau$ we have $\left[\hat{\mathcal{S}}_{\tau},\mathcal{O}\right]=0$. In the quadratic boundary term induced canonical transformation case, $\hat{\zeta}^n$ is one trivial example.
Subsequently,
\begin{equation}
\begin{aligned}
    \sandwitch{\Psi}{\mathcal{O}_{\zeta}\left(\hat{\zeta}_{\vk}(\tau_0),\hat{\pi}_{\vk}(\tau_0)\right)}{\Psi} \neq \sandwitch{\Psi^{(b)}}{\mathcal{O}_{\zeta}\left(\hat{\zeta}_{\vk}^{(b)}(\tau_0),\hat{\pi}_{\vk}^{(b)}(\tau_0)\right)}{\Psi^{(b)}}\hs,\\
    \sandwitch{\Psi}{\mathcal{O}_{a}\left(\ak(\tau_0),\admk(\tau_0)\right)}{\Psi} \neq \sandwitch{\Psi^{(b)}}{\mathcal{O}_{a}\left(\bk(\tau_0),\bdmk(\tau_0)\right)}{\Psi^{(b)}}\hs,\\
\end{aligned}
\end{equation}
and therefore one cannot get a consistent result for operators constructed by canonical variables or creation and annihilation operators. For the case of GKMR operators which are constructed with~(\ref{equ: single mode variables}), the expectation value of the Bell operator will also change when considering the quadratic boundary terms. As a simple analog, We can consider these $\hat{x}_{\vk}$ and $\hat{p}_{\vk}$ are just corresponding to another choice of $\mathcal{D}$ and in the next section, which we will carefully calculate.

Before we proceed to the next part, we may also consider a situation in which the canonical transformation is independent of time. In this case, we do not have $\mathcal{D}=\mathcal{D}^{(b)}$ since the canonical transformation at initial time does not go to identity. So the conclusion remains for the operator constructed by creation and annihilation operators while for the one constructed by canonical variables, the consistency recovers. 

The effect of the boundary term can be understood as the acceleration of the squeezing process. By calculating the squeezing parameter, we can easily demonstrate this idea. Consider the extra quadratic boundary term $\mathcal{L}_{\mathrm{bd}}\supset-\partial_{\tau}\left(9a^3HM_{pl}^2\zeta^2\right)$. Under an appropriate normalization, we get the Mukhanov-Sasaki variable~\cite{mukhanov:1981,Kodama:1984ziu} $\{y\equiv z\zeta, p_y\equiv y'-\frac{z'}{z}y\}$ which are suitable canonical variables to calculate the squeezing parameter, where prime is denoted to derivative to conformal time and $z\equiv aM_{pl}\sqrt{2\epsilon}$. Here we give the action in terms of the Mukhanov-Sasaki variable for reference:
\begin{equation}
    S=\frac{1}{2} \int d\tau \int d^3 x \left[ (y')^2-(\partial_i y)^2+\frac{z''}{z}y^2-\partial_{\tau}(\frac{z'}{z}y^2)+\partial_{\tau}(\frac{9}{\epsilon\tau}y^2) \right]\hs,
\label{equ: total action in ms variable}
\end{equation}
notice that the boundary term $\partial_{\tau}(\frac{z'}{z}y^2)$ naturally emerge when changing the variable from the original action~(\ref{equ:zeta action}).
In the remaining part of this section, we will use the same symbols in Fig.~\ref{fig:canonical transformation map} but $\{\hat{\zeta},\hat{\pi}\}$ are substituted by $\{\hat{y},\hat{p}_y \}$. Then we have
\begin{equation}
\begin{aligned}
    \mathcal{D}\begin{pmatrix}
\hat{y}_{\vk} \\
\hat{p}_{y,\vk}
\end{pmatrix}
&=\begin{pmatrix}
\sqrt{\frac{1}{2k}}(\ak+\admk)\\
-i\sqrt{\frac{k}{2}}(\ak-\admk)
\end{pmatrix}\hs,\\
\hat{y}_{\vk}(\tau)=\sqrt{\frac{1}{2k}}(\ak(\tau)+\admk(\tau))&=zf_k(\tau)\ak(\tau_0)+ zf_k^*(\tau)\admk(\tau_0)\hs,\\
\hat{p}_{y,\vk}(\tau)=-i\sqrt{\frac{k}{2}}(\ak(\tau)-\admk(\tau))&= -\frac{i}{z}\left[g_k(\tau)\ak(\tau_0)- g_k^*(\tau)\admk(\tau_0)\right]\hs,
\label{eq:quantization}
\end{aligned}
\end{equation}
and the relation between $\ak(\tau), \admk(\tau)$ and $\ak(\tau_0), \admk(\tau_0)$ can be described by Bogoliubov transformation
\begin{equation}
\begin{aligned}
    \ak(\tau) = u_k(\tau)\ak(\tau_0) +v_k(\tau)\admk(\tau_0)\hs,\\
    \admk(\tau) = u^*_k(\tau)\admk(\tau_0) +v^*_k(\tau)\ak(\tau_0)\hs,
\label{eq:bogoliubov}
\end{aligned}
\end{equation}
Preservation of canonical commuting relations require
\begin{equation}
    \left|u_k(\tau)\right|^2-\left|v_k(\tau)\right|^2=1\hs,
\end{equation}
so we can parameterize the functions $u_k$ and $v_k$ in the following way
\begin{equation}
\begin{aligned}
    u_k(\tau) &= e^{-i\theta_k(\tau)} \cosh r_k (\tau)\hs,\\
    v_k(\tau) &= e^{i\left(\theta_k(\tau)+2\varphi_k(\tau)\right)}\sinh r_k(\tau)\hs.
\end{aligned}
\end{equation}
where $r_k$ is the squeezing parameter we want to calculate, $\varphi_k$ is the squeezing angle and $\theta_k$ is the phase.

Take (\ref{eq:bogoliubov}) into (\ref{eq:quantization}) and compare with the coefficient. We can get
\begin{equation}
\begin{aligned}
    u_k(\tau) =\sqrt{\frac{kz^2}{2}}f_k +\sqrt{\frac{1}{2kz^2}}g_k\hs,\\
    v_k(\tau) = \sqrt{\frac{kz^2}{2}}f^*_k -\sqrt{\frac{1}{2kz^2}}g^*_k\hs,
\end{aligned}    
\end{equation}
and therefore the squeezing parameter can be calculated by
\begin{equation}
    \sinh^2 r_k = |v_k|^2= \left|\sqrt{\frac{kz^2}{2}}f_k -\sqrt{\frac{1}{2kz^2}}g_k\right|^2\hs,
\end{equation}
Take~(\ref{equ: mode of spectator}) into it we have\footnote{Here are two remarks. First, $u_k$ and $v_k$ can be directly solved from Heisenberg equations. Second, $\sinh^2 r_k$ equals to the average particle number $n_k$ of the squeezing state and a direct calculation can show this.}
\begin{equation}
    \sinh^2r_k = \left(\frac{1}{2k\tau}\right)^2\hs.
\label{equ: squeeze parameter with bd1}
\end{equation}

Subsequently, for the case with the extra boundary term, the particle number is defined as $\hat{n}^{(b)}_k = \bdk(\tau_0)\bk(\tau_0)$ so the result is just replacing $g_k$ by $g_k^{(b)}$ which is given by
\begin{equation}
\begin{aligned}
    p_y&=y'-\frac{z'}{z}y +\frac{9}{\epsilon \tau}y\hs,
\\
    \frac{g_k}{z} 
    &=i\sqrt{\frac{k}{2}}e^{-ik\tau}\left[-1-\frac{1}{{\epsilon }}\frac{9}{k^2\tau^2}\left(1+ik\tau\right)\right]\hs,    
\end{aligned}
\end{equation}
so we can obtain a different squeezing parameter:
\begin{equation}
    \sinh^2 r_k^{(b)} =\frac{(k\tau) ^2 (\epsilon +9)^2+81}{4(k\tau)^4 \epsilon ^2}\hs,
\end{equation}
Near $N=\log(\frac{aH}{k})=\log(\frac{1}{-k\tau})=0$, due to $\epsilon\ll1$, the squeezing parameter is much larger when considering the quadratic boundary term. Since the larger squeezing parameter leads to a larger violation of Bell inequalities~\cite{Martin:2017zxs}, we now come to the conclusion that the quadratic boundary term can affect the violation of Bell inequalities by accelerating the squeezing process.

\subsection{Inconsistency without extra quadratic boundary term}

From the above discussion, we know the definition of conjugate momentum can affect the squeezing parameter, and the change of the conjugate momentum can be done by adding a total time derivative term. In the above subsection, we obtain the squeezing parameter $r_k$ for a massless scalar field in quasi-de Sitter background~(\ref{equ: squeeze parameter with bd1}). However, in some literature~\cite{Kanno:2017dci,Kanno:2018cuk,Kanno:2019gqw,Kanno:2021vwu}, they obtain another value. The different definitions of conjugate momentum can give a clear explanation for it. 

For example in~\cite{Kanno:2017dci}. They introduce a truncation $\tau_r > 0$ which divides the inflation era and radiation-dominated era. Then the scale factor changes as
\begin{equation}
    a(\eta)=
    \left\{
    \begin{array}{lc}
         -\frac{1}{H(\tau-\tau_r)}, \quad\mathrm{for} \quad -\infty<\tau<\tau_r,  \\
          \frac{\tau}{H\tau_r^2}, \quad\quad\quad\hspace{2pt}\mathrm{for} \quad \tau_r<\tau.
    \end{array}
    \right.
\end{equation}
Now the mode functions $f_k$ of the Mukhanov-Sasaki variable are governed by different equations of motion and the solutions have the following expression
\begin{equation}
\begin{aligned}
    f_k^{\mathrm{in}} = \frac{1}{\sqrt{2k}}\left(1-\frac{i}{k(\tau-2\tau_r)}\right)e^{-ik(\tau-2\tau_r)},\quad&\mathrm{for} \quad -\infty<\tau<\tau_r,\\
    f_k^{\mathrm{out}} = \frac{1}{\sqrt{2k}}e^{-ik\tau},\quad &\mathrm{for} \quad \tau_r<\tau.    
\end{aligned}
\end{equation}
The first line is consistent with the first line of~(\ref{equ: mode of spectator}) up to a total phase factor after substituting $\tau\rightarrow\tau-2\tau_r$. Subsequently, the squeeze parameter is given by:
\begin{equation}
    \sinh^2r_k=\left|(f_k^{\mathrm{*out}},f_k^{\mathrm{in}})|_{\tau=\tau_r}\right|^2=\left(\frac{1}{2k^2\tau_r^2}\right)^2\hs,
\end{equation}
where $(f,g)\equiv i\{f^*g'-gf^{*'}\}$ is the Klein-Gordon inner product.

Even though it seems a different setup that they introduce a truncation $\tau_r$ for inflation, the equation of motion and the mode function remain the same after substituting $\tau\rightarrow\tau-2\tau_r$ and the vacuum states given by $\tau\rightarrow -\infty$ and $\tau>\tau_r$ respectively are physically the same since the mode functions are both $f_k=\frac{1}{\sqrt{2k}}e^{-ik\tau}$. Consequently, the squeeze parameter at $\tau=\tau_r$ should be given by~(\ref{equ: squeeze parameter with bd1}) (Note that $f_k$ and $g_k$ are now the mode function of Mukhanov-Sasaki variable, $z^2$ in the formula should be removed.)
\begin{equation}
    \sinh^2r_k(\tau_r)=\left.\left(\frac{1}{2k(\tau-2\tau_r)}\right)^2\right|_{\tau=\tau_r}=\left(\frac{1}{2k\tau_r}\right)^2\hs.
\label{equ:squeezing of y'}
\end{equation}
To explain this inconsistency, the boundary term $-\partial_{\tau}\left(\frac{z'}{z}y^2\right)$ in~(\ref{equ: total action in ms variable})
plays the key role.

The resolution of the inconsistency is repeating the calculation without the boundary term in~(\ref{equ: total action in ms variable}).
\begin{equation}
\begin{aligned}
    \tilde{p}_{y,\mathbf{k}}&=y'_{\mathbf{k}}\hs,\\
    \tilde{g}_k&=if'_k=i\sqrt{\frac{k}{2}}e^{-ik\tau}\left(\frac{1+ik\tau}{k^2\tau^2}-1\right)\hs,\\
    \sinh^2\tilde{r}_k &= \left|\left(\sqrt{\frac{k}{2}}f_k -\sqrt{\frac{1}{2k}}\tilde{g}_k\right)\right|^2 = \left(\frac{1}{2k^2\tau^2}\right)^2,
\end{aligned}
\end{equation}
which is consistent with~(\ref{equ:squeezing of y'}) and the reason for the inconsistency is the change of canonical variables from $\{y,p_{y}\equiv y'-\frac{z'}{z}y\}$ to $\{y,\tilde{p}_{y}\equiv y'\}$.

At the end of this section, we would like to emphasize our opinion regarding the selection of vacuum states and quadratic boundary terms. The fact that boundary terms can be seen as a canonical transformation which can affect the choice of vacuum, implies that we cannot arbitrarily choose boundary terms. Meanwhile, we indeed have a unique choice of boundary term fixed by the Gibbons-Hawking-York term in the action of gravity \cite{Sou:2022nsd,Ning:2023ybc}. Therefore, we believe that the action of a complete theory should include the information about the selection of the vacuum state, and correct calculations should incorporate quadratic boundary terms. Nonetheless, for reference purposes, we will discuss the cases with and without considering the quadratic boundary terms separately.

\section{The Expectation value of the Bell operator with decoherence}
\label{sec: expectation value}
For the cubic boundary interactions written in terms of system $\xi$ and environment $\mathcal{E}$, there are four types of terms: $\mathcal{E}\mathcal{E}\mathcal{E}$, $\xi\xi\xi$, $\mathcal{E}\mathcal{E}\xi$ and $\mathcal{E}\xi \xi$. The first one contributes a cubic pure phase $e^{i\int \mathcal{E}^3}$ so the tracing $\langle \exp\left(-i \int\mathcal{E}\mathcal{E}\mathcal{E}\right)\cdots\exp\left(i \int\mathcal{E}\mathcal{E}\mathcal{E}\right)\rangle_\mathcal{E}$ is trivial because the omitted part, represented by $\cdots$, usually commutes with this phase term and thus does not contribute to $\hat{\rho}_R$. The second one can contribute to the Bell inequality by a slow-roll unsuppressed cubic pure phase of $\xi$ to $\hat{\rho}_R$. However, it is negligible compared to the quadratic term (discussed in Section \ref{sec: effect of quadratic}), and it is also small compared to the last two terms involving environment modes since much less system modes are included due to momentum conservation. The last two terms involving interactions between the system and environment cause decoherence, which is expected to obstruct the Bell violation~\cite{Martin:2017zxs}. Therefore, for the cubic order, we expect that the leading contribution to the Bell inequality is caused by the dominated decoherence effect with the $\mathcal{E}\mathcal{E}\xi$ type interaction.

In this section, we calculate the expectation value of the Bell operator constructed by GKMR operators with considering decoherence, and we give the expression of GKMR operators in the field basis and discuss the effect of canonical transformation induced by quadratic boundary terms on the GKMR operators.

\subsection{GKMR operators in the field basis}
To calculate the expectation value of the Bell operator with wave functional, it is convenient to express GKMR operators in the field basis.
\begin{equation}
    \operatorname{Tr}(\hat{\rho}_R\hat{\mathcal{B}})=\int D\zeta D\tilde{\zeta}\sandwitch{\zeta}{\hat{\rho}_R}{\tilde{\zeta}}\sandwitch{\tilde{\zeta}}{\hat{\mathcal{B}}}{\zeta}=\int D\zeta D\tilde{\zeta} \rho_R[\zeta,\tilde{\zeta}]\mathcal{B}[\zeta,\tilde{\zeta}]\hspace{2pt},
\label{equ:expectation2}
\end{equation}
so we need to calculate $\sandwitch{\tilde{\zeta}}{\mathcal{B}}{\zeta}$. Because the Bell operator is made up of GKMR operators, now we calculate the matrix elements of GKMR operators in the field basis.

Firstly, we can simplify the GKMR operators in~(\ref{Sx}) and~(\ref{Sz})
\begin{equation}
\begin{aligned}
    \hat{S_z}(\mathbf{k}) &=-\int_{-\infty}^{\infty}dx_{\vk}\ket{x_{\vk}}\bra{-x_{\vk}}\hspace{2pt},\\
    \hat{S_x}(\mathbf{k}) &=\int_{0}^{\infty}dx_{\vk}(\ket{x_{\vk}}\bra{x_{\vk}}-\ket{-x_{\vk}}\bra{-x_{\vk}})\hspace{2pt}, 
\label{equ:reduced GKMR}
\end{aligned}
\end{equation}
from which we know the matrix elements are composed of functions with the form $\braket{\zeta_{\vk},\zeta_{-\vk}|x_{\vk},y_{-\vk}}$.  Recall the definitions of field operator and position operator and their corresponding conjugate momentum
\begin{equation}
\begin{aligned}
    \hat{\zeta}_{\vk}=\frac{\ak+\admk}{\sqrt{2z^2k}},\quad
    \hat{\pi}_{\vk}=-i\sqrt{\frac{z^2 k}{2}}(\ak-\admk)\hspace{2pt},\\
    \hat{x}_{\vk}=\frac{1}{\sqrt{2k}}(\ak+\adk),\quad
    \hat{p}_{\vk}=-i\sqrt{\frac{k}{2}}(\ak-\adk)\hspace{2pt}.    
\end{aligned}
\end{equation}
Recall $z\equiv aM_{pl}\sqrt{2\epsilon}$, the transition relations are given by
\begin{equation}
\begin{aligned}
    z \hat{\zeta}_{\vk}=\frac{1}{2}\left[\hat{x}_{\vk}+\hat{x}_{-\vk}+\frac{i}{k}(\hat{p}_{\vk}-\hat{p}_{-\vk})\right]\hspace{2pt},\\
    \frac{\hat{\pi}_{\vk}}{z} = \frac{1}{2i}\left[k(\hat{x}_{\vk}-\hat{x}_{-\vk})+i(\hat{p}_{\vk}+\hat{p}_{-\vk})\right]\hspace{2pt},
\end{aligned}
\end{equation}
and notice they satisfy the following two relations
\begin{align}
    z\hat{\zeta}_{\vk}+z\hat{\zeta}_{-\vk} = \hat{x}_{\vk}+\hat{x}_{-\vk}\hspace{2pt},  \label{equ:transition relation 1}\\
    z(\hat{\zeta}_{\vk}-\hat{\zeta}_{-\vk}) = \frac{i}{k}(\hat{p}_{\vk}-\hat{p}_{-\vk})\hspace{2pt}.\label{equ:transition relation 2}
\end{align}
Through these two relations, we can easily obtain the transition function
\begin{equation}
    \braket{x_{\vk},y_{-\vk}|\zeta_{\vk},\zeta_{-\vk}}=\braket{x_{\vk},y_{-\vk}|\zeta_{k1},\zeta_{k2}}=\sqrt{\frac{z^2k}{2\pi}}\delta\left(z\zeta_{k1}-\frac{x_{\vk}+y_{-\vk}}{2}\right)\exp\left(izk\zeta_{k2}(x_{\vk}-y_{-\vk})\right)\hspace{2pt}.
\end{equation}
where we define two real variables $\zeta_{k1}$ and $\zeta_{k2}$ for convenience
\begin{equation}
\begin{aligned}
    \zeta_{\vk}\equiv\zeta_{k1}+i\zeta_{k2}\hspace{2pt},\\
    \zeta_{-\vk}\equiv\zeta_{k1}-i\zeta_{k2} \hspace{2pt}.
\end{aligned}
\end{equation}

Then we can calculate the matrix elements of GKMR operators. In our case we only concern about $\hat{S}_z(\mathbf{k})\hat{S}_z(\mathbf{-k}), \hat{S}_x(\mathbf{k})\hat{S}_z(\mathbf{-k}), \hat{S}_z(\mathbf{k})\hat{S}_x(\mathbf{-k}), \hat{S}_x(\mathbf{k})\hat{S}_x(\mathbf{-k})$.  
By a quite straightforward calculation, we can obtain the matrix elements of these operators which are given by
\begin{equation}
\begin{aligned}
\braket{\tilde{\zeta}|\hat{S}_z(\mathbf{k})\hat{S}_x(-\mathbf{k})|\zeta}&=\frac{kz^2}{\pi}\exp\left(2ikz^2(\tilde{\zeta}_{k2}\zeta_{k1}-\zeta_{k2}\tilde{\zeta}_{k1})\right)\left(\theta(-\zeta_{k1}-\tilde{\zeta}_{k1})-\theta(\zeta_{k1}+\tilde{\zeta}_{k1})\right)\hspace{2pt},\\
\braket{\tilde{\zeta}|\hat{S}_x(\mathbf{k})\hat{S}_z(-\mathbf{k})|\zeta} &=\frac{kz^2}{\pi}\exp\left(-2ikz^2(\tilde{\zeta}_{k2}\zeta_{k1}-\zeta_{k2}\tilde{\zeta}_{k1})\right)\left(\theta(-\zeta_{k1}-\tilde{\zeta}_{k1})-\theta(\zeta_{k1}+\tilde{\zeta}_{k1})\right)\hspace{2pt},\\
\braket{\tilde{\zeta}|\hat{S}_z(\mathbf{k})\hat{S}_z(-\mathbf{k})|\zeta}&=\delta(\zeta_{k1}+\tilde{\zeta}_{k1})\delta(\zeta_{k2}+\tilde{\zeta}_{k2})\hspace{2pt},\\
\braket{\tilde{\zeta}|\hat{S}_x(\mathbf{k})\hat{S}_x(-\mathbf{k})|\zeta}&=\frac{4 k z^2}{\pi } \int_{0}^{\infty}dx_{\vk}\int_0^{\infty}dy_{-\vk} \left[\delta \left(x_{\vk}+y_{-\vk}-2 \zeta_{k1} z\right) \delta (x_{\vk}+y_{-\vk}-2 \tilde{\zeta}_{k1} z) \right.\\
&\left.\exp \left(i k z (\zeta_{k2}-\tilde{\zeta}_{k2}) (x_{\vk}-y_{-\vk})\right)+\left(\zeta,\tilde{\zeta}\to-\zeta,-\tilde{\zeta}\right)\right] \\
&-\delta(\zeta_{k1}-\tilde{\zeta}_{k1})\delta(\zeta_{k2}-\tilde{\zeta}_{k2})\hspace{2pt},
\label{equ:matrix elements}
\end{aligned}
\end{equation}
where $\theta(x)$ is the Heaviside function. 
\subsection{GKMR operators under canonical transformation}
The wave functional corresponding to the new vacuum is given by~(\ref{equ:new vacuum}). Then we need to construct the new Bell operator based on 
\begin{align}
    \hat{\zeta}^{(b)}_{\vk}=\frac{\bk+\bdmk}{\sqrt{2z^2k}}=\hat{\zeta}_{\bf k},\quad
    \hat{\pi}^{(b)}_{\vk}=\sqrt{\frac{z^2 k}{2}}\left[-i(\bk-\bdmk)\hspace{2pt}+\lambda_k\left(\bk+\bdmk\right)\right] \ , \label{equ:new_canonical_quant}
\end{align}
where $\lambda_k=\frac{9}{\epsilon k\tau_0}$ coming from the quadratic boundary term (\ref{equ:quadratic_bd_term}). Similar to the free theory, we can flip the comoving momentum $-{\vk} \to {\vk}$ in (\ref{equ:new_canonical_quant}) to define the 'position' operator and its conjugate momentum:
\begin{align}
    \hat{x}^{(b)}_{\vk}=\frac{1}{\sqrt{2k}}(\bk+\bdk),\quad
    \hat{p}^{(b)}_{\vk}=\sqrt{\frac{k}{2}}\left[-i(\bk-\bdk)\hspace{2pt}+\lambda_k(\bk+\bdk)\right]\hspace{2pt}.    \label{equ:new_x_p}
\end{align}
We define the eigenstate of the new 'position' operator
\begin{equation}
  \hat{x}^{(b)}_{\vk}\ket{x^{(b)}_{\vk}}=x_{\vk}\ket{x^{(b)}_{\vk}}\hs,
\end{equation}
where the superscript $(b)$ is not written for eigenvalues to avoid complicated expressions in the following calculation.
The following relation is useful to derive the transformation of basis $ \braket{x^{(b)}_{\vk},y^{(b)}_{-\vk}|\zeta_{\vk},\zeta_{-\vk}}$
\begin{align}
   \hat{x}^{(b)}_{\vk}+\hat{x}^{(b)}_{-\vk}&= z\hat{\zeta}_{\vk}+z\hat{\zeta}_{-\vk} \hspace{2pt},  \label{equ:transition relation 1b}\\
    \hat{p}^{(b)}_{\vk}-\hat{p}^{(b)}_{-\vk}&=-ikz\left[\hat{\zeta}_{\vk}-\hat{\zeta}_{-\vk}+i\frac{\lambda_k}{z}\left(\hat{x}_{\vk}^{(b)}-\hat{x}_{-{\vk}}^{(b)}\right)\right]\hspace{2pt},\label{equ:transition relation 2b}
\end{align}
with which the transformation obtains an additional Gaussian phase factor from the boundary term
\begin{align}
    \braket{x^{(b)}_{\vk},y^{(b)}_{-\vk}|\zeta_{\vk},\zeta_{-\vk}}&=\braket{x^{(b)}_{\vk},y^{(b)}_{-\vk}|\zeta_{k1},\zeta_{k2}} \nonumber \\
    &=\sqrt{\frac{z^2k}{2\pi}}\delta\left(z\zeta_{k1}-\frac{x_{\vk}+y_{-\vk}}{2}\right)e^{izk\zeta_{k2}(x_{\vk}-y_{-\vk})+i\frac{\lambda_k k}{4}\left(x_{\vk}-y_{-\vk}\right)^2}\hspace{2pt}. \label{equ:new_transformation_basis}
\end{align}
It is easy to notice the extra total phase term $\exp{\left[i\frac{\lambda_k k}{4}\left(x_{\vk}-y_{-\vk}\right)^2\right]}$ will be naturally canceled in $\braket{\tilde{\zeta}|\hat{S}^{(b)}_x(\mathbf{k})\hat{S}^{(b)}_x(-\mathbf{k})|\zeta}$ and $\braket{\tilde{\zeta}|\hat{S}^{(b)}_z(\mathbf{k})\hat{S}^{(b)}_z(-\mathbf{k})|\zeta}$. For the $\braket{\tilde{\zeta}|\hat{S}^{(b)}_x(\mathbf{k})\hat{S}^{(b)}_z(-\mathbf{k})|\zeta}$ and $\braket{\tilde{\zeta}|\hat{S}^{(b)}_z(\mathbf{k})\hat{S}^{(b)}_x(-\mathbf{k})|\zeta}$, the integrands are still odd so their expectation values vanish. 

    The definition of the Bell operator is unchanged, and we emphasize that the Bell operator $\hat{\mathcal{B}}^{(b)}$ constructed by flipping $-{\bf k}\to {\bf k}$ (\ref{equ:new_x_p}) should not be confused with the one from canonical transformation having the form $\hat{\mathcal{S}}\hat{\mathcal{B}}\hat{\mathcal{S}}^{-1}$. However, the quantum state is changed by the quadratic boundary term, as seen in the wave functional (\ref{equ: dynamic term with bd}). So we get different expectation values of the Bell operator if we consider the quadratic boundary term. Recall the discussion in last section, and the condition for consistent result under canonical transformation is $\hat{\mathcal{S}}_{\tau}\hat{\mathcal{B}}\hat{\mathcal{S}}^{-1}_{\tau}=\hat{\mathcal{B}}^{(b)}$. However, the above calculation shows this condition is not satisfied since $\hat{\mathcal{B}}=\hat{\mathcal{B}}^{(b)}$ and $\hat{\mathcal{B}}$ does not commute with $\hat{\mathcal{S}}_{\tau}$.

\subsection{The Expectation value of the Bell operator}
Now we can calculate the expectation value of the Bell operator with respect to the density matrix~(\ref{equ:reduce density matrix}). Firstly, we simply demonstrate the process of proving $\braket{\hat{S}_x(\mathbf{k})\hat{S}_z(-\mathbf{k})}=\braket{\hat{S}_z(\mathbf{k})\hat{S}_x(-\mathbf{k})}=0$, since the calculation is simple and direct. We only need to prove that the integrands are odd under the transformation $\zeta_{k1}, \tilde{\zeta}_{k1}, \zeta_{k2}, \tilde{\zeta}_{k2} \rightarrow -\zeta_{k1}, -\tilde{\zeta}_{k1}, -\zeta_{k2}, -\tilde{\zeta}_{k2}$. The integrand is a product of two parts, the reduced density matrix~(\ref{equ:reduce density matrix}) which is even under the above transformation and the matrix elements $\braket{\tilde{\zeta}|\hat{S}_x(\mathbf{k})\hat{S}_z(-\mathbf{k})|\zeta}$ and $\braket{\tilde{\zeta}|\hat{S}_z(\mathbf{k})\hat{S}_x(-\mathbf{k})|\zeta} $ which we expect to be odd to complete the proof. Notice the Heaviside function parts in the expression of the matrix elements~(\ref{equ:matrix elements}), these parts change the sign under the above transformation and the rest parts are even. Therefore, these two terms vanish under the integration and do not contribute to the Bell Inequality. Finally, we obtain~(\ref{equ:Bell operator}).

Then we only need to calculate the value of $\braket{\hat{S}_x(\mathbf{k})\hat{S}_x(-\mathbf{k})}$ and $\braket{\hat{S}_z(\mathbf{k})\hat{S}_z(-\mathbf{k})}$. The reduced density matrix $\hat{\rho}_\mathrm{R}$ is given in~(\ref{equ:reduce density matrix}) and the matrix elements of $\hat{S}_x(\mathbf{k})\hat{S}_x(-\mathbf{k})$ and $\hat{S}_z(\mathbf{k})\hat{S}_z(-\mathbf{k})$ are given in~(\ref{equ:matrix elements}). Take all these expressions into~(\ref{equ:expectation2}). Then we can obtain the expression of $\braket{\hat{S}_z(\mathbf{k})\hat{S}_z(-\mathbf{k})}$
\begin{equation}
\begin{aligned}
&\braket{\hat{S}_z(\mathbf{k})\hat{S}_z(-\mathbf{k})} \\
&=\int_{-\infty}^{\infty}d\zeta_{k1} d\zeta_{k2} d\tilde{\zeta}_{k1} d\tilde{\zeta}_{k2} \frac{2\operatorname{Re}A_{\xi}(k,\tau)}{\pi}\exp\left(-A_{\xi}(k,\tau)\left(\zeta_{k1}^2+\zeta_{k2}^2\right)-A^*_{\xi}(k,\tau)\left(\tilde{\zeta}_{k1}^2+\tilde{\zeta}_{k2}^2\right)\right)\\
    &\exp\left(-\tilde{\Gamma}\left(\left(\zeta_{k2}-\tilde{\zeta}_{k2}\right)^2+\left(\zeta_{k1}-\tilde{\zeta}_{k1}\right)^2\right)\right)\delta\left(\zeta_{k1}+\tilde{\zeta}_{k1}\right)\delta\left(\zeta_{k2}+\tilde{\zeta}_{k2}\right)\\
    &=\frac{2\operatorname{Re}A_{\xi}(k,\tau)}{\pi}\int_{-\infty}^{\infty}d\zeta_{k1} d\zeta_{k2}\exp\left(-\left(A_{\xi}(k,\tau)+A_{\xi}^*(k,\tau)\right)(\zeta_{k1}^2+\zeta_{k2}^2)\right)\exp\left(-4\tilde{\Gamma}\left(\zeta_{k2}^2+\zeta_{k1}^2\right)\right)\\
    &=\frac{\operatorname{Re}A_{\xi}(k,\tau)}{\operatorname{Re}A_{\xi}(k,\tau)+2\tilde{\Gamma}}\hspace{2pt}.
\label{equ:szsz}
\end{aligned}
\end{equation}
This value has a clear physical meaning that is the purity of the density matrix as mentioned in \cite{Martin:2022kph}.

For $\braket{\hat{S}_x(\mathbf{k})\hat{S}_x(-\mathbf{k})}$ which is more difficult, we need some tricks. The matrix elements of $\braket{\hat{S}_x(\mathbf{k})\hat{S}_x(-\mathbf{k})}$ contains a complicated integral and a unit element, as shown in~(\ref{equ:matrix elements}). The expectation value of a unit operator is $1$, so we only need to deal with the first part of~(\ref{equ:matrix elements}) denoted by $\braket{\hat{S}_x(\mathbf{k})\hat{S}_x(-\mathbf{k})}^{(2)}$. First, we integrate $\zeta_{k1}$ and $\tilde{\zeta}_{k1}$. Since there are two $\delta$ functions, the result is just a substitution $\zeta_{k1}, \zeta_{k2} \rightarrow \frac{x_{\vk}+y_{-\vk}}{2z}$. Notice that $\int_{-\infty}^{\infty} dx \delta(\alpha x)=1/|\alpha|$. Then we obtain
\begin{equation}
\begin{aligned}
    &\braket{\hat{S}_x(\mathbf{k})\hat{S}_x(-\mathbf{k})}^{(2)}=\frac{4k\operatorname{Re}A_{\xi}(k,\tau)}{\pi^2}\int_0^{\infty}dx \int_0^{\infty}dy   \exp\left(-2\operatorname{Re}A_{\xi}(k,\tau)\left(\frac{x+y}{2z}\right)^2\right) \\
  &\int_{-\infty}^{\infty}d \zeta_{k2} \int_{-\infty}^{\infty}d\tilde{\zeta}_{k2}
  \cos \left(k z (\zeta_{k2}-\tilde{\zeta}_{k2}) (x-y)\right)\exp\left(-A_{\xi}(k,\tau)\zeta_{k2}^2-A_{\xi}^*(k,\tau)\tilde{\zeta}_{k2}^2-\tilde{\Gamma}(\zeta_{k2}-\tilde{\zeta}_{k2})^2\right) \hspace{2pt},   
\end{aligned}
\end{equation}
It is convenient to make a variable substitution
\begin{equation}
    \lambda=\zeta_{k2}+\tilde{\zeta}_{k2}\quad, \quad \eta = \zeta_{k2} - \tilde{\zeta}_{k2}\hspace{2pt},
\end{equation}
Finally, we obtain
\begin{equation}
\begin{aligned}
    \braket{\hat{S}_x(\mathbf{k})\hat{S}_x(-\mathbf{k})}^{(2)}=&\frac{4k\operatorname{Re}A_{\xi}(k,\tau)}{\pi^2}\int_0^{\infty}dx \int_0^{\infty}dy   \exp\left(-2\operatorname{Re}A_{\xi}(k,\tau)\left(\frac{x+y}{2z}\right)^2\right) \\
  &\frac{1}{2}\int_{-\infty}^{\infty}d\eta
  \cos (k z (x-y)\eta)\exp\left(-\frac{\operatorname{Re}A_{\xi}(k,\tau)}{2}\eta^2-\tilde{\Gamma}\eta^2\right)\\
  &\int_{-\infty}^{\infty}d \lambda \exp\left(-\frac{\operatorname{Re}A_{\xi}(k,\tau)}{2}\lambda^2-\frac{A_{\xi}(k,\tau)-A_{\xi}^*(k,\tau)}{2}\lambda\eta\right)\\
=&1+\frac{2}{\pi } \arctan\left(\frac{k^2z^4-|A_{\xi}(k,\tau)|^2-2\operatorname{Re}A_{\xi}(k,\tau)\tilde{\Gamma}}{2 kz^2 \sqrt{|A_{\xi}(k,\tau)|^2+2\operatorname{Re}A_{\xi}(k,\tau)\tilde{\Gamma}}}\right)
\hspace{2pt},
\end{aligned}
\end{equation}
where we use the following Gaussian formulas
\begin{equation}
\begin{aligned}
    \int_{-\infty}^{\infty}dx \exp\left(-Axy-x^2\right)&=\exp\left(\frac{A^2y^2}{4}\right)\sqrt{\pi}\hspace{2pt},
\\
    \int_{-\infty}^{\infty}dx \cos\left(Ax\right)\exp\left(-x^2\right)&=\exp\left(-\frac{A^2}{4}\right)\sqrt{\pi}\hspace{2pt}.
\end{aligned}
\end{equation}
The total expectation of $\braket{\hat{S}_x(\mathbf{k})\hat{S}_x(-\mathbf{k})}=-1+\braket{\hat{S}_x(\mathbf{k})\hat{S}_x(-\mathbf{k})}^{(2)}$  is given by
\begin{equation}
    \braket{\hat{S}_x(\mathbf{k})\hat{S}_x(-\mathbf{k})}=\frac{2}{\pi } \arctan\left(\frac{k^2z^4-|A_{\xi}(k,\tau)|^2-2\operatorname{Re}A_{\xi}(k,\tau)\tilde{\Gamma}}{2 kz^2 \sqrt{|A_{\xi}(k,\tau)|^2+2\operatorname{Re}A_{\xi}(k,\tau)\tilde{\Gamma}}}\right)
\hspace{2pt},
\label{equ:sxsx}
\end{equation}
However, one may notice the result in \cite{Martin:2022kph} is inconsistent with ours if we change the phenomenological parameter from $\tilde{\Gamma}$ to the purity of the density matrix by $p=\frac{\operatorname{Re}A_{\xi}(k,\tau)}{\operatorname{Re}A_{\xi}(k,\tau)+2\tilde{\Gamma}}$. A simple explanation is that in their setup there is only one parameter, purity, and they modify the covariance matrix by dividing by the square root of the purity. However, there is no unique way to modify the covariance matrix while keeping the purity and in our case the two-point correlation function $\braket{\zeta^2}$ is unchanged. Therefore, we do not have the same result.

The results without decoherence can easily be obtained by letting $\tilde{\Gamma}=0$

\begin{align}
\braket{\hat{S}_z(\mathbf{k})\hat{S}_z(-\mathbf{k})}_{\tilde{\Gamma}=0}&=1\hspace{2pt},\\
\braket{\hat{S}_x(\mathbf{k})\hat{S}_x(-\mathbf{k})}_{\tilde{\Gamma}=0}&=\frac{2}{\pi } \arctan\left(\frac{k^2z^4-|A_{\xi}(k,\tau)|^2}{2 kz^2 |A_{\xi}(k,\tau)|}\right)
\hspace{2pt},
\end{align}
these results are consistent with~\cite{Martin:2017zxs}.

\section{Bell test curve with decoherence}
\label{sec: effect of decoherence}   
    In this section, we introduce the concept of ``Bell test curves'', give two examples of decoherence and plot their Bell test curves. The first is decoherence from gravitational nonlinearity for which we review the details in Section~\ref{sec:set up}. The second one is phenomenological. We consider the power-law decoherence rate and discuss the behavior of the Bell test curve in the superhorizon regime.
\subsection{Decoherence from nonlinearity of gravity}
\begin{figure}
    \centering
    \hspace{-2cm}
    \includegraphics[width=0.8\textwidth]{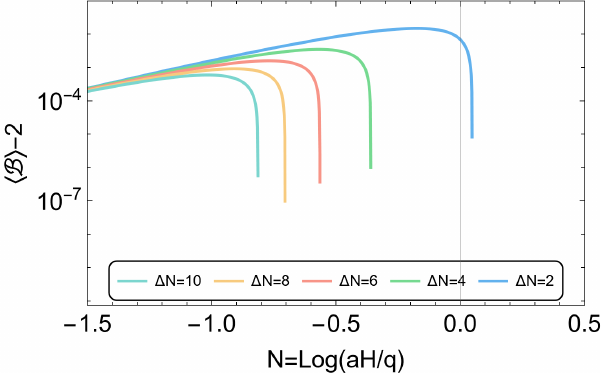}
    \caption{The violation of the Bell inequality for $\Gamma_{\mathrm{bd}}$.} 
    \label{fig:violation}
\end{figure}

We define $N\equiv\log(aH/q)$ the e-folds after crossing the horizon. Then from~(\ref{equ:dynamics term}), (\ref{equ:Bell operator}), (\ref{equ:szsz}) and (\ref{equ:sxsx}) we obtain the exact dependence of the expectation of Bell operator with variables $N$ in the case of ignoring the quadratic boundary term,
\begin{equation}
\braket{\hat{\mathcal{B}}}=2\sqrt{\left(\frac{1}{1+2 \Gamma (\exp (-2 N)+1) }\right)^2+\frac{4}{\pi^2}\arctan^2\left(\frac{1-\frac{2\Gamma\exp (-2 N)+1}{\exp (2 N)+1}}{2\sqrt{\frac{2\Gamma\exp (-2 N)+1}{\exp (2 N)+1}}}\right)}\hspace{2pt},
\label{equ:result}
\end{equation}
where we apply the observed values \cite{Planck:2018jri}
    \begin{equation}
        \Delta_\zeta^2 \approx 2.5 \times 10^{-9}\hspace{2pt}, \epsilon<0.006\hspace{2pt}, \eta \approx 0.03\hspace{2pt},
        \label{eq: observed values}
    \end{equation}
to evaluate the decoherence rate $\Gamma$.
\begin{figure}
    \centering
    \hspace{2.3cm}
    \includegraphics[width=0.8\textwidth]{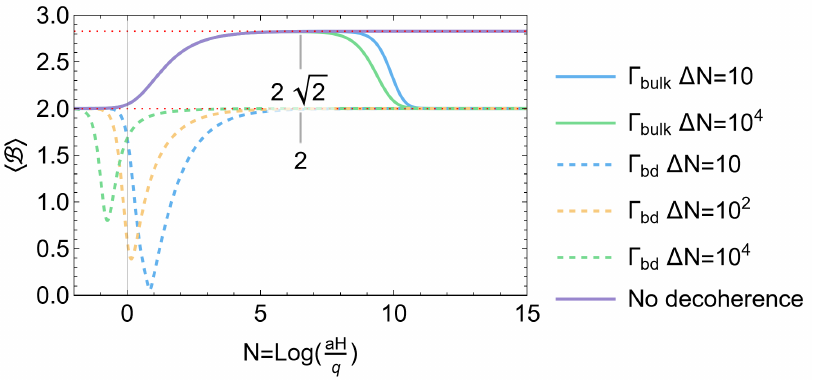}
    \caption{Expectation value of Bell operator $\operatorname{Tr}(\hat{\rho}_{\mathrm{R}}\hat{\mathcal{B}})$ for the decoherence induced by gravitational nonlinearity in bulk term and boundary term when the Bell operator is constructed by GKMR pseudo-spin operators as the function of e-folds elapsed after crossing the horizon $N$ with different values of renormalization parameter $\Delta N$. Dotted lines illustrate the classical and Cirel'son bound.}
    \label{fig:result}
\end{figure}
Fig.~\ref{fig:result} demonstrates the relation between the expectation value of the Bell operator and e-folds elapsed after crossing the horizon $N$ and the Bell test curve that we mentioned many times. It shows that, in terms of the expectation values of the Bell operator, the boundary term leads to faster decoherence compared to the bulk term. For decoherence caused by the boundary term, the modes around the horizon decohere rapidly, however, for the bulk term, it takes about 10 e-folds for the expectation value of the Bell operator to decrease below the classical upper limit after crossing the horizon. The behavior of the expectation value of the Bell operator varies for different $\Gamma$ and different $\Delta N$, indicating that we can distinguish between different sources of decoherence through cosmological Bell tests. Fig.~\ref{fig:violation} clearly shows that in the case of boundary term induced decoherence, the Bell inequality is still slightly violated. Higher $\Delta N$ has smaller largest violation and becomes decohered earlier. Since the violation is small, it is needed to examine the next leading contribution. In Appendix~\ref{app:higher loop}, we estimated the contribution from higher loop (i.e. the next leading order term in (\ref{eq:moments expansion})). According to the estimation, the results here are still reliable.

\begin{figure}
    \centering
    \hspace{2.1cm}
    \includegraphics[width=0.8\textwidth]{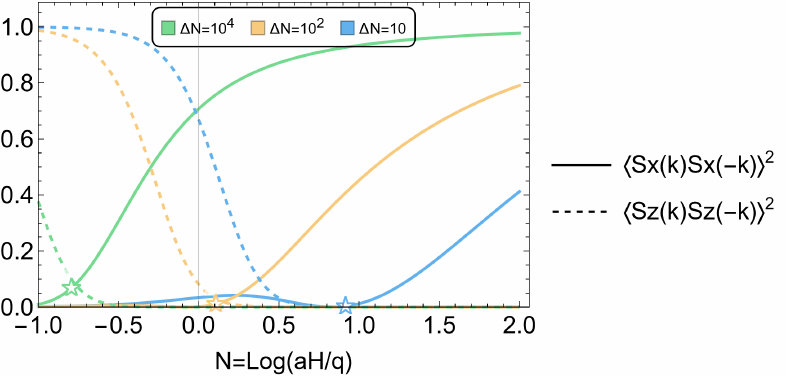}
    \caption{The value of $\braket{\hat{S}_x(\mathbf{k})\hat{S}_x(-\mathbf{k})}^2$~(\ref{equ:sxsx}) and $\braket{\hat{S}_z(\mathbf{k})\hat{S}_z(-\mathbf{k})}^2$~(\ref{equ:szsz}) for $\Gamma_{\mathrm{bd}}$. The star markers indicate the value of $N$ which minimized $\braket{\hat{\mathcal{B}}}$.} 
    \label{fig:feature}
\end{figure}

One important difference between the cases of the boundary and the bulk term is the existence of the local minimum for the expectation value of the Bell operator. For the boundary term, as shown in Fig.~\ref{fig:feature}, we notice that this local minimum occurs close to the point when $\braket{\hat{S}_z(\vk)\hat{S}_z(-\vk)}^2 = \braket{\hat{S}_x(\vk)\hat{S}_x(-\vk)}^2$.  Also, notice the value of $\braket{\hat{S}_z(\vk)\hat{S}_z(-\vk)}^2$ decreased by the increasing of $\Gamma$, while the value of $\braket{\hat{S}_x(\vk)\hat{S}_x(-\vk)}^2$ is increasing. Since $\Gamma_{\mathrm{bd}}$ increases much faster than $\Gamma_{\mathrm{bulk}}$, before $\braket{\hat{S}_x(\vk)\hat{S}_x(-\vk)}^2$ goes to near 1, $\braket{\hat{S}_z(\vk)\hat{S}_z(-\vk)}^2$ goes to 0 . Therefore there is a local minimum for the case of $\Gamma_{\mathrm{bd}}$.

\begin{figure}
    \centering
    \includegraphics[width=0.8\textwidth]{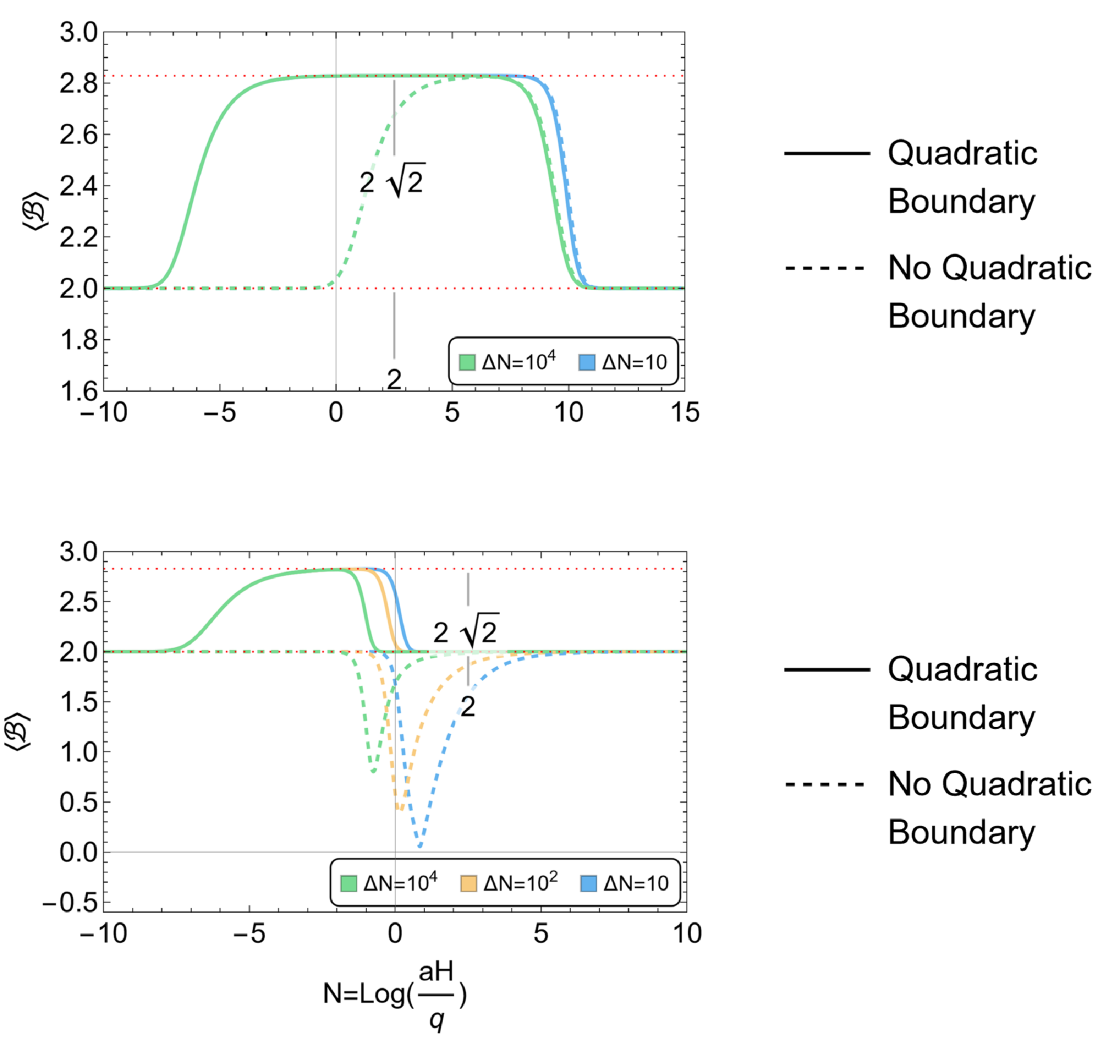}
    \caption{Expectation value of Bell operator $\operatorname{Tr}(\hat{\rho}_{\mathrm{R}}\hat{\mathcal{B}})$ when considering (i) upper panel: bulk term and (ii) lower panel: boundary term induced decoherence. The solid lines represent the case when considering the phase induced by the quadratic boundary term $A_{\zeta}^{bd}\equiv A_{\zeta}+18ia^3 H M_{pl}^2$ and the dashed lines represent the case without this extra term.}
    \label{fig:result2}
\end{figure}

Now we consider the contribution from the quadratic boundary term. Putting~(\ref{equ: dynamic term with bd}) into~(\ref{equ:Bell operator}), (\ref{equ:szsz}) and (\ref{equ:sxsx}), we obtain
\begin{equation}
\braket{\hat{\mathcal{B}}}^{(b)}=2\sqrt{\left(\frac{1}{1+2 \Gamma (e^{-2 N}+1) }\right)^2+\frac{4}{\pi^2}\arctan^2\left(\frac{1-\frac{(2 \Gamma  e^{-2 N} +1)\epsilon ^2+9 e^{2 N} (2 \epsilon +9)+81 e^{4 N}}{\left(e^{2 N}+1\right) \epsilon ^2}}{2\sqrt{\frac{(2 \Gamma  e^{-2 N} +1)\epsilon ^2+9 e^{2 N} (2 \epsilon +9)+81 e^{4 N}}{\left(e^{2 N}+1\right) \epsilon ^2}}}\right)}\hspace{2pt}.
\label{equ:result2}
\end{equation}
Fig.~\ref{fig:result2} 
shows the acceleration effect of the quadratic boundary term on the squeezing process. In the upper panel, the bulk term induced decoherence case, the time of reaching the maximal violation of Bell inequalities has been advanced. While the lower panel shows a different feature from the case without quadratic boundary term that before the decoherence the quantum state of the fluctuation has the maximal violation of Bell inequalities. However, the time region of restoring Bell inequalities remains the same in both cases.

In all the cases except the one without decoherence in the large $N$ limit (i.e. superhorizon limit), the expectation value of the Bell operator goes to near 2. Detailed calculation in the next example shows they are exactly under 2, so the Bell inequality is restored and we cannot prove the large scale structure originates from quantum fluctuation through the cosmological Bell test in the superhorizon regime.

\subsection{Power-law decoherence rate}
As shown in the above example, the Bell inequality is restored in the superhorizon regime, and so far we do not have a probe to measure the Bell operator during inflation. So, a natural question is whether we can extract information from the superhorizon behavior of the Bell test curve. We will give a simple answer in this part. 

We assume the decoherence rate $\Gamma$ is a polynomial of $\exp{(N)}$ and as $N\gg1$ we will also have $\Gamma\gg1$. So, we consider the following form of decoherence rate
\begin{equation}
    \Gamma_n \supset A_n\exp(nN)\hs,
\end{equation}
where $A_n>0$ and $n>0$. In general, we can only keep the leading order term. However, in some special cases as we will see later, the next leading order gives the main contribution and we will add the next order term back in that case.

Here we will also discuss $A_{\zeta}$ the quadratic coefficient of the free Gaussian state in two cases, given by~(\ref{equ:dynamics term}) without the quadratic boundary term and~(\ref{equ: dynamic term with bd}) with the quadratic boundary term respectively.

\paragraph{Without the quadratic boundary term}

The expectation value of the Bell operator is given by~(\ref{equ:result}) which contains $\braket{\hat{S}_z(\mathbf{k})\hat{S}_z(-\mathbf{k})}$ and $\braket{\hat{S}_x(\mathbf{k})\hat{S}_x(-\mathbf{k})}$ two parts. In the superhorizon limit, then we have
    \begin{equation}
        \braket{\hat{S}_z(\mathbf{k})\hat{S}_z(-\mathbf{k})}_{\mathrm{sup}} \approx \frac{1}{2\Gamma_n}=\frac{1}{2A_n}\exp{(-nN)}\hs,
    \end{equation}
    where the subscript $\mathrm{sup}$ means the superhorizon limit. However, for $\braket{\hat{S}_x(\mathbf{k})\hat{S}_x(-\mathbf{k})}$, the leading order term in the superhorizon limit cannot be simply determined. We first simplify it. Since $n$ is still undetermined, for leading order approximation we can only obtain 
    \begin{align}
        \braket{\hat{S}_x(\mathbf{k})\hat{S}_x(-\mathbf{k})} 
        &=\frac{2}{\pi}\arctan\left(\frac{1-\frac{2\Gamma\exp (-2 N)+1}{\exp (2 N)+1}}{2\sqrt{\frac{2\Gamma\exp (-2 N)+1}{\exp (2 N)+1}}}\right)\label{equ:original ESxSx}\\
        &\overset{\mathrm{sup}}\approx \frac{2}{\pi}\arctan{\left(\exp{(N)}\frac{1-2\Gamma_n\exp{(-4N)}}{2\sqrt{1+2\Gamma_n\exp{(-2N)}}}\right)}\hs.
    \end{align}
From the expression we can expect two possible critical points that are $n=2$ and $n=4$ and therefore splitting into three regions $n<2$, $2<n<4$ and $n>4$. We first discuss the case of $n<2$.

For $n<2$, in the superhorizon limit, the effect of decoherence can be ignored in this term and we have
\begin{equation}
    \braket{\hat{S}_x(\mathbf{k})\hat{S}_x(-\mathbf{k})}_{\mathrm{sup}, n<2} \approx \frac{2}{\pi}\arctan{\left(\frac{\exp{(N)}}{2}\right)}\approx1-\frac{4}{\pi}\exp(-N)\hs,
\end{equation}
where we use $\arctan{(1/x)}=\frac{\pi}{2}-x+O(x^3), x>0$. As $x$ goes to zero, we only keep the leading approximation. Then the expectation value of the Bell operator is given by
\begin{equation}
\begin{aligned}
    \braket{\hat{\mathcal{B}}}_{\mathrm{sup},n<2}&\approx2\sqrt{1+\frac{\exp{(-2nN)}}{4A_n^2}-\frac{8}{\pi}\exp(-N)}\\
    &\approx 2+\frac{\exp{(-2nN)}}{4A_n^2}-\frac{8}{\pi}\exp(-N)
    \hs,    \label{equ:B_sup_nless2}
\end{aligned}    
\end{equation}
Here we only care about the leading term so we need to compare $2n$ and $1$ which gives us another critical point $n=1/2$. This point means the main contribution to the expectation value of the Bell operator changes from $\braket{\hat{S}_z(\mathbf{k})\hat{S}_z(-\mathbf{k})}$ to $\braket{\hat{S}_x(\mathbf{k})\hat{S}_x(-\mathbf{k})}$. Therefore, we have 3 critical points and 4 regions. In principle, we do not require $n$ to be an integer, we can have $n<1/2$, then we have
\begin{equation}
\braket{\hat{\mathcal{B}}}_{\mathrm{sup},n<1/2}\approx 2+\frac{\exp{(-2nN)}}{4A_n^2}>2
    \hs.    
\end{equation}
So, in the $n<1/2$ region, we even still have the violation of Bell inequalities though very small. Other properties we care about are the slope and intercept of the logarithmic Bell test curve, $\log\left(\left|\braket{\hat{\mathcal{B}}}-2\right|\right)$ versus $N$.
\begin{equation}
    \log\left(\braket{\hat{\mathcal{B}}}_{\mathrm{sup}, n<1/2}-2\right)=-2nN-2\log(2A_n)\hs,
\end{equation}
the slope is the power of the leading order $-2n$ and the intercept is $-2\log(2A_n)$ from which we can extract all the information about the leading decoherence rate.

For $n=1/2$
\begin{equation}
\braket{\hat{\mathcal{B}}}_{\mathrm{sup}, n=1/2}\approx 2+\exp{(-N)}\left(\frac{1}{4A_{1/2}^2}-\frac{8}{\pi}\right)
    \hs,   
\end{equation}
the violation of Bell inequalities will remain in the superhorizon limit with large $A_{1/2}<\sqrt{\frac{\pi}{32}}$, while for $A_{1/2}>\sqrt{\frac{\pi}{32}}$, Bell inequalities restores. For the very special case $A_{1/2}=\sqrt{\frac{\pi}{32}}$, the next leading order of decoherence rate will also play an important role. Suppose the next leading order decoherence rate is $\Gamma_m = A_m\exp{(mN)}$ and $m<\frac{1}{2}$ ($A_m$ can be zero in the case there is only one term in decoherence rate). We need to carefully compare the order of the power:
\begin{equation}
\begin{aligned}
     \braket{\hat{S}_z(\mathbf{k})\hat{S}_z(-\mathbf{k})}_{\mathrm{sup}, n=1/2, A_{1/2}=\sqrt{\pi/32}}&=\frac{1}{1+\left(\Gamma_{1/2}+\Gamma_{m}\right)\left(1+\exp(-2N)\right)}\\
     &\approx\frac{1}{1+A_{1/2}\exp\left(\frac{1}{2}N\right)+A_m\exp(mN)+A_{1/2}\exp\left(-\frac{3}{2}N\right)}\hs,
\end{aligned}
\end{equation}
and
\begin{equation}
\begin{aligned}
    \braket{\hat{S}_x(\mathbf{k})\hat{S}_x(-\mathbf{k})}_{\mathrm{sup}, n=1/2, A_{1/2}=\sqrt{\pi/32}}=1-\frac{4}{\pi}\exp(-N)-\frac{4A_{1/2}}{\pi}\exp\left(-\frac{5}{2}N\right)\hs,
\end{aligned}
\end{equation}
since in the leading order, it does not contain $A_{1/2}$, and we do not need to contain $\Gamma_m$ to get the next leading order result. Therefore, if $A_m\neq0$ (i.e. the next leading term of decoherence rate exists), we have
\begin{equation}
\begin{aligned}
    \braket{\hat{S}_z(\mathbf{k})\hat{S}_z(-\mathbf{k})}_{\mathrm{sup}, n=1/2, A_{1/2}=\sqrt{\pi/32}} \approx \frac{1}{2\Gamma_{1/2}+2\Gamma_m}\approx\frac{1}{2\Gamma_{1/2}}\left(1-\frac{\Gamma_m}{\Gamma_{1/2}}\right)\hs,\\
    \braket{\hat{\mathcal{B}}}_{\mathrm{sup}, n=1/2, A_{1/2}=\sqrt{\pi/32}}\approx 2-\frac{\Gamma_{m}}{2\Gamma_{1/2}^3}=2-\frac{A_m}{2}\left(\frac{32}{\pi}\right)^\frac{3}{2}\exp{\left(\left(m-\frac{3}{2}\right)N\right)}\hs.  
\end{aligned}
\end{equation}
Since we do not require $A_m>0$ the Bell inequality can still be violated if $A_m<0$.
And if $A_m=0$
\begin{equation}
\begin{aligned}
    \braket{\hat{S}_z(\mathbf{k})\hat{S}_z(-\mathbf{k})}_{\mathrm{sup}, n=1/2, A_{1/2}=\sqrt{\pi/32}} \approx \frac{1}{1+2\Gamma_{1/2}}=\frac{1}{2\Gamma_{1/2}}\left(1-\frac{1}{2\Gamma_{1/2}}\right)\hs,\\
    \braket{\hat{\mathcal{B}}}_{\mathrm{sup}, n=1/2, A_{1/2}=\sqrt{\pi/32}}\approx 2-\frac{1}{4\Gamma_{1/2}^3}=2-\frac{1}{4}\left(\frac{32}{\pi}\right)^\frac{3}{2}\exp{\left(-\frac{3}{2}N\right)}\hs.  
\end{aligned}
\end{equation}

For $1/2<n<2$
\begin{equation}
\braket{\hat{\mathcal{B}}}_{\mathrm{sup},1/2<n<2}\approx 2-\frac{8}{\pi}\exp{(-N)}
    \hs,   
\end{equation}
In this region, no specific information from the decoherence is left. They are degenerate to the same slope $-1$ and same intercept $\log(\frac{8}{\pi})$.

For $n=2$
\begin{equation}
    \braket{\hat{S}_x(\mathbf{k})\hat{S}_x(-\mathbf{k})}_{\mathrm{sup}, n=2} \approx \frac{2}{\pi}\arctan{\left(\frac{\exp{(N)}}{2\sqrt{1+2A_2}}\right)}\hs,
\end{equation}
As mentioned before, for $n>1/2$, we can ignore $\braket{\hat{S}_z(\mathbf{k})\hat{S}_z(-\mathbf{k})}$ then we get
\begin{equation}
\begin{aligned}
    \braket{\hat{\mathcal{B}}}_{\mathrm{sup}, n=2}\approx 2-\frac{8}{\pi}\sqrt{1+2A_2}\exp{(-N)}
    \hs,
\end{aligned}
\end{equation}

For $2<n<4$
\begin{equation}
\begin{aligned}
    \braket{\hat{S}_x(\mathbf{k})\hat{S}_x(-\mathbf{k})}_{\mathrm{sup}, 2<n<4} &\approx \frac{2}{\pi}\arctan{\left(\frac{\exp{\left[(2-\frac{n}{2})N\right]}}{2\sqrt{2A_n}}\right)}\hs,\\
    \braket{\hat{\mathcal{B}}}_{\mathrm{sup}, 2<n<4}&\approx 2-\frac{8}{\pi}\sqrt{2A_n}\exp{\left[\left(\frac{n}{2}-2\right)N\right]}
    \hs.
\end{aligned}
\end{equation}

For $n=4$, if we only keep the leading term, we have
\begin{equation}
    \braket{\hat{S}_x(\mathbf{k})\hat{S}_x(-\mathbf{k})}_{\mathrm{sup}, n=4} \approx \frac{2}{\pi}\arctan\left(\frac{1-2A_4}{2\sqrt{2A_4}}\right)\hs,
\end{equation}
which is a constant. This is the only case where the expectation value of the Bell operator does not go to $2$, and to get the slope we should consider the contribution from the next leading order. We should also define the logarithmic Bell test curve correspondingly, $\log\left(\left|\braket{\hat{\mathcal{B}}}-\frac{4}{\pi}\arctan\frac{1-2A_4}{2\sqrt{2A_4}}\right|\right)$ versus $N$. Suppose the next leading order term of decoherence rate is $\Gamma_m=A_m\exp{(mN)}$ and $m<4$. We have
\begin{equation}
\begin{aligned}
    &\braket{\hat{S}_z(\mathbf{k})\hat{S}_z(-\mathbf{k})}_{\mathrm{sup},n=4}\approx\frac{1}{2A_4}\exp(-4N)\hs,\\
     &\braket{\hat{S}_x(\mathbf{k})\hat{S}_x(-\mathbf{k})}_{\mathrm{sup},n=4}\\
     &\approx \frac{2}{\pi}\arctan\left(\frac{1-2A_4}{2\sqrt{2A_4}} +\frac{1+2A_4}{4\sqrt{2A_4}}\left(1-\frac{1}{2A_4}-\frac{A_m}{A_4}\exp\left[(m-2)N\right]\right)\exp(-2N)\right)
\end{aligned}
\end{equation}
so $\braket{\hat{S}_z(\mathbf{k})\hat{S}_z(-\mathbf{k})}$ can still be ignored and $\braket{\hat{\mathcal{B}}}\approx2\braket{\hat{S}_x(\mathbf{k})\hat{S}_x(-\mathbf{k})}$.\\
For $2<m<4$
\begin{equation}
\begin{aligned}
    \braket{\hat{S}_x(\mathbf{k})\hat{S}_x(-\mathbf{k})}_{\mathrm{sup},n=4,2<m<4}
    &\approx\frac{2}{\pi}\arctan\left(\frac{1-2A_4}{2\sqrt{2A_4}}\right)-\frac{2A_m\sqrt{2A_4}}{(1+2A_4)A_4\pi}\exp\left[(m-4)N\right]\hs.
\end{aligned}
\end{equation}
For $m<2$ or $A_m=0$
\begin{equation}
\begin{aligned}
    \braket{\hat{S}_x(\mathbf{k})\hat{S}_x(-\mathbf{k})}_{\mathrm{sup},n=4,m<2}
    &\approx\frac{2}{\pi}\arctan\left(\frac{1-2A_4}{2\sqrt{2A_4}}\right)+\frac{2\sqrt{2A_4}}{(1+2A_4)\pi}\left(1-\frac{1}{2A_4}\right)\exp(-2N)\hs.
\end{aligned}
\end{equation}
For $m=2$ and $A_m\neq A_4-\frac{1}{2}$
\begin{equation}
\begin{aligned}
    \braket{\hat{S}_x(\mathbf{k})\hat{S}_x(-\mathbf{k})}_{\mathrm{sup},n=4,m=2}
    &\approx\frac{2}{\pi}\arctan\left(\frac{1-2A_4}{2\sqrt{2A_4}}\right)+\frac{2\sqrt{2A_4}}{(1+2A_4)\pi}\left(1-\frac{1}{2A_4}-\frac{A_m}{A_4}\right)\exp(-2N)\hs.
\end{aligned}
\end{equation}
For $m=2$ and $A_m = A_4-\frac{1}{2}$, one can directly verify with~(\ref{equ:original ESxSx}) that
\begin{equation}
    \braket{\hat{S}_x(\mathbf{k})\hat{S}_x(-\mathbf{k})}_{n=4,m=2}=\arctan\left(\frac{1-2 A_4}{2 \sqrt{2} \sqrt{A_4}}\right)\hs,
\end{equation}
which is a constant, to get its slope and intercept information, we need to consider the third-order decoherence rate (if it exists),
\begin{equation}
    \Gamma = A_4\exp(4N) + \left(A_4-\frac{1}{2}\right)\exp(2N) + A_p \exp(pN)\hs,
\end{equation}
where $0<p<2$. Then we have
\begin{align}
        &\braket{\hat{S}_x(\mathbf{k})\hat{S}_x(-\mathbf{k})}_{\mathrm{sup},n=4,m=2}=\arctan\left(\frac{1-2 A_4 \left(\frac{A_p \exp(pN)}{A_4 \left(\exp(4N)+\exp(2N)\right)}+1\right)}{2 \sqrt{2} \sqrt{A_4 \left(\frac{A_p \exp(pN)}{A_4 \left(\exp(4N)+\exp(2N)\right)}+1\right)}}\right)\nonumber\\
        &\approx\frac{2}{\pi}\arctan\left(\frac{1-2 A_4}{2 \sqrt{2} \sqrt{A_4}}\right)-\frac{2 \sqrt{2} A_p \exp[(p-4)N]}{\sqrt{A_4} \pi(2 A_4+ 1)}\hs.
\end{align}

For $n>4$
\begin{equation}
\begin{aligned}
    \braket{\hat{S}_x(\mathbf{k})\hat{S}_x(-\mathbf{k})}_{\mathrm{sup}, n>4} &\approx -\frac{2}{\pi}\arctan{\left(\frac{\sqrt{2A_n}}{2}\exp{\left[\left(\frac{n}{2}-2\right)N\right]}\right)}\hs,\\
    \braket{\hat{\mathcal{B}}}_{\mathrm{sup}, n>4}&\approx 2-\frac{8}{\pi\sqrt{2A_n}}\exp{\left[\left(2-\frac{n}{2}\right)N\right]}
    \hs.
\end{aligned}
\end{equation}

\paragraph{With the quadratic boundary term}

    The expectation value of the Bell operators is given by~(\ref{equ:result2}) and $\braket{\hat{S}_z(\mathbf{k})\hat{S}_z(-\mathbf{k})}$ is unchanged compared to the case without the quadratic boundary term. The significant differences appear in $\braket{\hat{S}_x(\mathbf{k})\hat{S}_x(-\mathbf{k})}^{(b)}$. So we again simplify it to see possible critical points

\begin{equation}
        \braket{\hat{S}_x(\mathbf{k})\hat{S}_x(-\mathbf{k})}^{(b)}_{\mathrm{sup}} \approx -\frac{2}{\pi}\arctan{\left(\frac{\exp{(N)}}{2}\sqrt{2 \Gamma_n  \exp{(-6 N)} +\frac{81}{\epsilon^2} }\right)}\hs,
    \end{equation}
from the expression, we know we still have a critical point at $n=\frac{1}{2}$ and another critical point at $n=6$. 

For $n<6$
\begin{equation}
        \braket{\hat{S}_x(\mathbf{k})\hat{S}_x(-\mathbf{k})}^{(b)}_{\mathrm{sup}, n<6} \approx -\frac{2}{\pi}\arctan{\left(\frac{9\exp{(N)}}{2\epsilon}\right)}\approx-1+\frac{4\epsilon}{9\pi}\exp{(-N)}\hs.
    \end{equation}
Take it into the expression of the expectation value of the Bell operator,
\begin{equation}
\begin{aligned}
    \braket{\hat{\mathcal{B}}}^{(b)}_{\mathrm{sup}, n<6}&\approx2\sqrt{1+\frac{\exp{(-2nN)}}{4A_n^2}-\frac{8\epsilon}{9\pi}\exp(-N)}\\
    &\approx 2+\frac{\exp{(-2nN)}}{4A_n^2}-\frac{8\epsilon}{9\pi}\exp(-N)
    \hs,    
\end{aligned}    
\end{equation}
The structure is very similar to the case without the quadratic boundary term (\ref{equ:B_sup_nless2}), and the last negative term becomes slow-roll suppressed, leading to a larger expectation value.

For $n<1/2$
\begin{equation}
\begin{aligned}
    \braket{\hat{\mathcal{B}}}^{(b)}_{\mathrm{sup}, n<1/2}&\approx2+\frac{\exp{(-2nN)}}{4A_n^2}\hs,
\end{aligned}
\end{equation}
we still have the violation of Bell Inequalities.

For $n=1/2$
\begin{equation}
\begin{aligned}
    \braket{\hat{\mathcal{B}}}^{(b)}_{\mathrm{sup}, n=1/2}\approx 2+\left(\frac{1}{4A_{1/2}^2}-\frac{8\epsilon}{9\pi}\right)\exp(-N)
    \hs,    
\end{aligned}    
\end{equation}
if $\frac{1}{4A_{1/2}^2}>\frac{8\epsilon}{9\pi}$ the violation of Bell inequalities will remain, while for $\frac{1}{4A_{1/2}^2}<\frac{8\epsilon}{9\pi}$, the Bell inequality restores. For the special case $A_{1/2}=\sqrt{\frac{9\pi}{32\epsilon}}$, we still consider a next leading term in decoherence rate $\Gamma_m = A_m \exp(mN)$ and $m<1/2$. Now the next leading order of $\braket{\hat{S}_x(\mathbf{k})\hat{S}_x(-\mathbf{k})}^{(b)}_{\mathrm{sup}, n=1/2}$ is $\exp\left(-3N\right)$ which is smaller that the case of without the quadratic boundary term which is $\exp\left(-\frac{5}{2}N\right)$. Since $\braket{\hat{S}_x(\mathbf{k})\hat{S}_x(-\mathbf{k})}$ is the same as the former case, we get almost the same result (the only different is the critical $A_{1/2}$).
\begin{equation}
\begin{aligned}
    \braket{\hat{\mathcal{B}}}^{(b)}_{\mathrm{sup}, n=1/2, A_{1/2}=\sqrt{\frac{9\pi}{32\epsilon}}}\approx 2-\frac{\Gamma_{m}}{2\Gamma_{1/2}^3}=2-\frac{A_m}{2}\left(\frac{32\epsilon}{9\pi}\right)^\frac{3}{2}\exp{\left(\left(m-\frac{3}{2}\right)N\right)}\hs,  
\end{aligned}
\end{equation}
if $A_m=0$
\begin{equation}
\begin{aligned}
    \braket{\hat{\mathcal{B}}}^{(b)}_{\mathrm{sup}, n=1/2, A_{1/2}=\sqrt{\frac{9\pi}{32\epsilon}}}\approx 2-\frac{1}{4\Gamma_{1/2}^3}=2-\frac{1}{4}\left(\frac{32\epsilon}{9\pi}\right)^\frac{3}{2}\exp{\left(-\frac{3}{2}N\right)}\hs.  
\end{aligned}
\end{equation}

For $1/2<n<6$
\begin{equation}
\begin{aligned}
    \braket{\hat{\mathcal{B}}}^{(b)}_{\mathrm{sup}, n<6}\approx 2-\frac{8\epsilon}{9\pi}\exp(-N)
    \hs.    
\end{aligned}    
\end{equation}

For $n=6$
\begin{equation}
\begin{aligned}
    \braket{\hat{S}_x(\mathbf{k})\hat{S}_x(-\mathbf{k})}^{(b)}_{\mathrm{sup}, n=6} \approx -\frac{2}{\pi}\arctan{\left(\frac{\exp{(N)}}{2}\sqrt{2 A_6   +\frac{81}{\epsilon^2} }\right)}\hs,\\
    \braket{\hat{\mathcal{B}}}^{(b)}_{\mathrm{sup}, n=6}\approx 2-\frac{8\epsilon}{\pi\sqrt{2 A_6 \epsilon^2   +81 }}\exp(-N)
    \hs. 
\end{aligned}
\end{equation}

For $n>6$

\begin{equation}
\begin{aligned}
    \braket{\hat{S}_x(\mathbf{k})\hat{S}_x(-\mathbf{k})}^{(b)}_{\mathrm{sup}, n>6} \approx -\frac{2}{\pi}\arctan{\left(\frac{\exp{(N)}}{2}\sqrt{2 A_n \exp{\left[(n-6)N\right]} }\right)}\hs,\\
    \braket{\hat{\mathcal{B}}}^{(b)}_{\mathrm{sup}, n>6}\approx 2-\frac{8}{\pi\sqrt{2 A_n }}\exp\left[\left(2-\frac{n}{2}\right)N\right]
    \hs. 
\end{aligned}
\end{equation}

\begin{table}[htbp]
  \centering
  \caption{The slope and intercept of the logarithmic Bell test curve for the case without the quadratic boundary term.}
    \begin{tabular}{cccc}
    \toprule
    \toprule
    Leading Term (and higher order term) & \multicolumn{3}{c}{Without Quadratic Boundary Term} \\
    \cmidrule{2-4}\makecell{$\Gamma_{n,(m,p)}=A_ne^{-nN}(+A_me^{-mN}$\\$+A_pe^{-pN})$} & Slope & Intercept & \multicolumn{1}{c}{\makecell{Violation of\\ Bell Inequalities}} \\
    \midrule
    $n<1/2$ & $-2n$ & $-2\log(2A_n)$ & Yes \\
    $n=1/2,A_{1/2}\neq\sqrt{\pi/32}$ & $-1$ & $\log\left(\frac{1}{4A_{1/2}^2}-\frac{8}{\pi}\right)$  & \makecell{Yes$(A^2_{1/2}<\frac{\pi}{32})$\\No$(A^2_{1/2}>\frac{\pi}{32})$} \\
    $n=1/2,A_{1/2}=\sqrt{\pi/32},A_m\neq0$ & $m-3/2$  & $\log\left(\frac{|A_m|}{2}\left(\frac{32}{\pi}\right)^\frac{3}{2}\right)$  & \makecell{Yes$(A_m<0)$\\No$(A_m>0)$} \\
    $n=1/2,A_{1/2}=\sqrt{\pi/32},A_m=0$ & $-3/2$  & $\log\left(\frac{1}{4}\left(\frac{32}{\pi}\right)^\frac{3}{2}\right)$  & No \\
    $1/2<n<2$ & $-1$  & $\log(\frac{8}{\pi})$  & No \\
    $n=2$ & $-1$  & $\log\left(\frac{8}{\pi}\sqrt{1+2A_2}\right)$  & No \\
    $2<n<4$ & $n/2-2$  & $\log\left(\frac{8}{\pi\sqrt{2A_n}}\right)$  & No \\
    $n=4,2<m<4$ & $m-4$  & $\log\left(\frac{4|A_m|\sqrt{2A_4}}{(1+2A_4)A_4\pi}\right)$  & No \\
    $n=4,m=2,A_m\neq A_4-\frac{1}{2}$ & $-2$  & $\log\left(\frac{4\sqrt{2A_4}}{(1+2A_4)\pi}\left(1-\frac{1}{2A_4}-\frac{A_m}{A_4}\right)\right)$  & No \\
    $n=4,m=2,A_m= A_4-\frac{1}{2},0<p<2$ & $p-4$  & $\log\left(\frac{4 \sqrt{2} A_p }{\sqrt{A_4} \pi(2 A_4+ 1)}\right)$  & No \\
    $n=4,m=2,A_m= A_4-\frac{1}{2}, A_p=0$ &\multicolumn{3}{c} {The expectation value of the Bell operator is a Constant}\\
    $n=4,m<2$ or $A_m=0$ & $-2$ & $\log\left(\frac{4\sqrt{2A_4}}{(1+2A_4)\pi}\left(1-\frac{1}{2A_4}\right)\right)$ & No \\
    $n>4$ & $2-n/2$ & $\log
    \left(\frac{8}{\pi\sqrt{2A_n}}\right)$ & No \\
    \bottomrule
    \bottomrule
    \end{tabular}
  \label{tab: without quadratic bd term}
\end{table}

\begin{table}[htbp]
  \centering
  \caption{The slope and intercept of the logarithmic Bell test curve for the case with the quadratic boundary term.}
    \begin{tabular}{cccc}
    \toprule
    \toprule
    Leading Term (and next leading term) & \multicolumn{3}{c}{With Quadratic Boundary Term} \\
    \cmidrule{2-4}$\Gamma_{n,(m)}=A_ne^{-nN}(+A_me^{-mN})$ & Slope & Intercept & \multicolumn{1}{c}{\makecell{Violation of\\ Bell Inequalities}} \\
    \midrule
    $n<1/2$ & $-2n$ & $-2\log(2A_n)$ & Yes \\
    $n=1/2,A_{1/2}\neq\sqrt{9\pi/32\epsilon}$ & $-1$ & $\log\left(\frac{1}{4A_{1/2}^2}-\frac{8\epsilon}{9\pi}\right)$  & \makecell{Yes$(A^2_{1/2}<\frac{9\pi}{32\epsilon})$\\No$(A^2_{1/2}>\frac{9\pi}{32\epsilon})$} \\
    $n=1/2,A_{1/2}=\sqrt{9\pi/32\epsilon},A_m\neq0$ & $m-\frac{3}{2}$  & $\log\left(\frac{|A_m|}{2}\left(\frac{32\epsilon}{9\pi}\right)^\frac{3}{2}\right)$  & \makecell{Yes$(A_m<0)$\\No$(A_m>0)$} \\
    $n=1/2,A_{1/2}=\sqrt{9\pi/32\epsilon},A_m=0$ & $-\frac{3}{2}$  & $\log\left(\frac{1}{4}\left(\frac{32\epsilon}{9\pi}\right)^\frac{3}{2}\right)$  & No \\
    $1/2<n<6$ & $-1$  & $\log(\frac{8\epsilon}{9\pi})$  & No \\
    $n=6$ & $-1$  & $\log\left(\frac{8\epsilon}{\pi\sqrt{2 A_6 \epsilon^2   +81 }}\right)$  & No \\
    $n>6$ & $2-\frac{n}{2}$ & $\log
    \left(\frac{8}{\pi\sqrt{2A_n}}\right)$ & No \\
    \bottomrule
    \bottomrule
    \end{tabular}
  \label{tab: with quadratic bd term}
\end{table}

\begin{figure}
    \centering
    \hspace{-0.8cm}
    \includegraphics[width=0.8\textwidth]{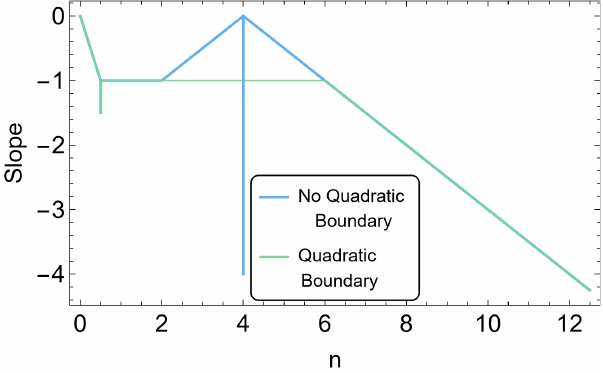}
    \caption{The slope of the logarithmic Bell test curve as the function of the power of the leading term of the decoherence rate $n$. The blue/green line corresponds to the case that the wave functional includes/excludes the contribution from the quadratic boundary term.}
    \label{fig:Slope versus power}
\end{figure}

In the Table.~\ref{tab: without quadratic bd term} and Table.~\ref{tab: with quadratic bd term}, we summarize the result above, and Fig.~\ref{fig:Slope versus power} demonstrates the relation between the slope of the logarithmic Bell test curve and the power of leading decoherence terms. What we first learn from the calculation is that for slow decoherence we may still have the violation of Bell inequalities, but this is unusual since the power is not an integer. Another property is there is a range that we can get almost nothing about the decoherence source from the slope and intercept of the logarithmic Bell test curve: for the case without quadratic boundary term it is $1/2<n<2$, and for the case with the quadratic boundary term case it is $1/2<n<6$. However, fortunately, the decoherence from the gravitational nonlinearity of the boundary term does not lie in these ranges, whereas for the bulk-term decoherence, this happens only when the quadratic boundary term is neglected. Although we also discussed some very special cases, it is less likely that it will be the case. Finally, we can answer the question at the beginning of this part, that we can get the leading power and the coefficient of the decoherence rate from the slope and intercept of the logarithmic Bell test curve in the superhorizon limit respectively, despite degeneracy in some special cases.

\section{Conclusion and outlook}
\label{sec:conclusion}
We investigated the effects of quantum decoherence in cosmological Bell tests. We used the Bell operator constructed by the GKMR operator and quantitatively determined the relationship between the expectation value of the Bell operator and the number of e-folds before/after the mode exiting the horizon with decoherence considered. The results can be demonstrated by Bell test curves.

We discussed the effect of the quadratic boundary term from the full action on the Bell inequality. We found that the definition of GKMR operators is unchanged under the canonical transformation induced by quadratic boundary terms. However, the expectation value of the Bell operator changes because of the mismatching of the canonical transformation between the quantum state and observables. The effect of the quadratic boundary term can also be understood as an acceleration of the squeezing process.

We provided two examples for different focuses. The first example is quantum decoherence induced by gravitational nonlinearity, which does not require the introduction of other fields or interactions and provides a lower limit for the total decoherence effect. Therefore, this example gives us the upper bound of the range of the violation of Bell inequalities for directly determining the quantum origin of the cosmic structure formation. We calculated the results with neglecting and considering the quadratic boundary term respectively. In the former case, we compared the Bell test curve with decoherence from the bulk term and boundary term. For the bulk term, before the complete decoherence at around 10 e-folds, there is a several e-folds long range of maximal violation. But for the boundary term case, the decoherence happens faster and the Bell inequality is only slightly violated. In the latter case, we found the maximal violation of Bell inequalities for both the bulk and boundary term induced decoherence case due to the effect of the acceleration of the squeezing process.

The second example is based on the question of how to extract the information of the decoherence from the superhorizon limit of the Bell test curve. We choose the dominated power of $\exp(N)=aH/q$ and the coefficient of the decoherence rate in the superhorizon regime as phenomenological parameters. We find that the slope and the intercept of the logarithmic Bell test curve can carry the information of these two parameters. It may provide us with a new way to test decoherence and help us understand the quantum-to-classical transition during the early universe.

In this paper, we only considered the leading boundary terms on superhorizon scales (\ref{equ:boundary term}) characterized by the largest positive order of scale factor $a(t)^3$, thus estimating the fastest changes of the Bell inequality. On the other hand, some sub-dominated terms proportional to $a(t)$ with spatial derivatives can appear on boundary even at the quadratic order
\begin{align}
\mathcal{L}^{\rm bd}_{\rm int}\supset M_{pl}^2\partial_\tau\left[\frac{a\left(\partial_i\zeta\right)^2}{H}+\frac{a\zeta\left(\partial_i\zeta\right)^2}{H}\right]+\mathcal{O}\left(a^{-1}\partial^4,\epsilon,\eta\right) \ ,
\end{align}
collectively included in the term $M_{pl}^2\frac{a^3e^{3\zeta}}{2H}{^{(3)}R}$, with ${^{(3)}R}$ the three dimensional Ricci scalar \cite{Ning:2023ybc}. Such terms can be significant on sub-horizon scales while setting initial conditions for the canonical transformation and tracing out UV environment modes, but we expect that their contributions to late-time observables grow much slower than the terms proportional to $a(t)^3$ in (\ref{equ:boundary term}). More rigorous and detailed treatment deserves future study. 

\section{Acknowledgment}
We thank Ding Ding, Chen-Te Ma, Amaury Micheli and Fumiya Sano for the helpful discussion during the workshop "Gravity and Cosmology 2024". JW and YW were supported in part by the National Key R\&D Program of China (2021YFC2203100). CMS was supported by NSFC under Grant No. 12275146, the Dushi Program of Tsinghua University and the Shuimu Tsinghua Scholar Program.

\bibliographystyle{utphys.bst}
\bibliography{decoherence}
\appendix
\section{Estimation of higher loop correction}
\label{app:higher loop}
For no quadratic boundary term case, boundary term induced decoherence will lead to slight violation of Bell inequality Fig.~\ref{fig:violation}. It is worthy to check whether higher loop correction (i.e. higher order term in (\ref{eq:moments expansion})) has significant correction on it.

\begin{figure}
	\centering
	\includegraphics[width=0.6\textwidth]{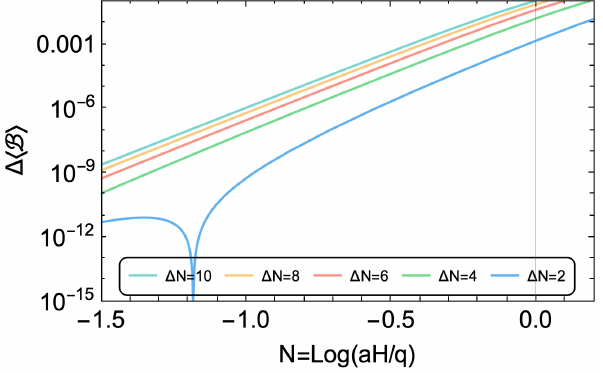}
	\caption{The growth of $\Delta \langle \mathcal{B}\rangle$. The size of the quartic term is set to $\kappa=\tilde{\Gamma}_{\mathrm{bd}}^2$. \label{fig:errorBell}}
\end{figure}

\begin{figure}
	\centering
	\includegraphics[width=0.8\textwidth]{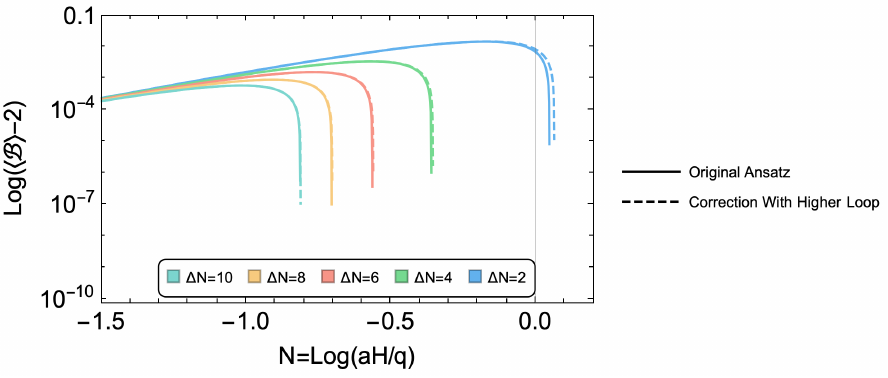}
	\caption{The correction of the Bell violation. The size of the quartic term is set to $\kappa=\tilde{\Gamma}_{\mathrm{bd}}^2$. \label{fig:errorBell2}}
\end{figure}

The correction of $\rho_{R}$ is controlled by
\begin{equation}
	\begin{aligned}[b]
		\rho_{R}(\xi,\tilde{\xi})
		&=\Psi_G(\xi)\Psi_G^*(\tilde{\xi})\exp\left[- \mathcal{F}^2 \frac{\langle\mathcal{E}^4\rangle_c}{2!} \left(\xi-\tilde{\xi}\right)^2+\mathcal{F}^4\frac{\langle \mathcal{E}^8\rangle_c}{4!}\left(\xi-\tilde{\xi}\right)^4+\cdots\right] \\
		&\approx \Psi_G(\xi)\Psi_G^*(\tilde{\xi})\exp\left[- \mathcal{F}^2\frac{\langle\mathcal{E}^4\rangle_c}{2!} \left(\xi-\tilde{\xi}\right)^2\right]\left[1+\mathcal{F}^4\frac{\langle \mathcal{E}^8\rangle_c}{4!}\left(\xi-\tilde{\xi}\right)^4\right] \, \
	\end{aligned}
\end{equation}
where we use a simplified symbol $\mathcal{F}^2\equiv|\mathcal{F}^2|\approx(\mathrm{Im}\mathcal{F})^2$ and omit the subscript of the mode ${\bf q}$. The second approximation is valid as long as the perturbativity holds. Now we introduce the parameters $\tilde\Gamma\sim \mathcal{F}^2 \mathcal{O}\left(\langle \mathcal{E}^4\rangle_c\right)$ and $\kappa\sim \mathcal{F}^4\mathcal{O}\left(\langle \mathcal{E}^8\rangle_c\right)$ to represent the sizes of the quadratic and quartic term
\begin{equation}
	\begin{aligned}[b]
		\rho_R(\xi,\tilde{\xi};\tilde\Gamma,\kappa)&\approx \Psi_G(\xi)\Psi_G^*(\tilde{\xi}) \exp\left(-\tilde\Gamma(\xi-\tilde{\xi})^2\right)\left[1+\kappa (\xi-\tilde{\xi})^4\right] \\
		&=\left(1+\kappa \frac{\partial^2}{\partial \tilde\Gamma^2}\right)\rho_R(\xi,\tilde{\xi};\tilde\Gamma,0) \, \ ,
	\end{aligned}
\end{equation}
where clearly $\rho_R(\xi,\tilde{\xi};\tilde\Gamma,0)$ is the ansatz used in the paper. The accuracy of the expectation of the Bell operator is thus controlled by 
\begin{equation}
	\begin{aligned}[b]
		\Delta \langle \mathcal{B}\rangle
		&={\rm Tr}\left[\mathcal{B}\left(\rho_R(\xi,\tilde{\xi};\tilde\Gamma,\kappa)-\rho_R(\xi,\tilde{\xi};\tilde\Gamma,0)\right)\right]\\
		&=\kappa \frac{\partial^2}{\partial \tilde\Gamma^2}{\rm Tr}\left(\mathcal{B}\rho_R(\xi,\tilde{\xi};\tilde\Gamma,0)\right) \\ 
            &=\kappa \frac{16\pi^4 \Delta^4_{\zeta}}{q^6} \frac{\partial^2}{\partial \Gamma^2}{\rm Tr}\left(\mathcal{B}\rho_R(\xi,\tilde{\xi};\tilde\Gamma,0)\right) \,  \ .
	\end{aligned}
\end{equation}
In Fig. \ref{fig:errorBell}, we set the size of the quartic term $\kappa=\tilde{\Gamma}_{\mathrm{bd}}^2$ for estimation. So the prefactor is just $\Gamma_{\mathrm{bd}}^2$. Notice the maximal violation is got before decoherence ($\Gamma\simeq1$), so we expect the correction is small. 
From the figure, we can see that the error of Bell inequality is from $\mathcal{O}(10^{-6})$ to $\mathcal{O}(10^{-3})$ around the maximal violation point. A direct comparison in Fig.~\ref{fig:errorBell2} shows that the ansatz is reliable even in the case that the violation is small.

It is noteworthy that we do not exactly calculate the expression of $\kappa$ since it is complicated for higher loop correction. A detailed calculation for it is beyond this paper. 
\end{document}